\documentclass[pra,twocolumn,amsmath,amssymb,superscriptaddress]{revtex4-1}

\usepackage{epsfig,amsmath}
\usepackage{subfigure}
\usepackage{graphicx}
\usepackage{dcolumn}
\usepackage{stmaryrd}
\usepackage{mathrsfs}
\usepackage{pifont}
\usepackage{amsthm}
\usepackage{amssymb}
\usepackage{bm}
\usepackage{latexsym}
\usepackage[colorlinks=true,linkcolor=blue,citecolor=blue]{hyperref}
\usepackage{color}
\usepackage{epstopdf}

\begin{document}

\title{Chiral current in Floquet cavity-magnonics}

\author{Shi-fan Qi}
\affiliation{School of Physics, Zhejiang University, Hangzhou 310027, Zhejiang, China}

\author{Jun Jing}
\email{jingjun@zju.edu.cn}
\affiliation{School of Physics, Zhejiang University, Hangzhou 310027, Zhejiang, China}

\date{\today}

\begin{abstract}
Floquet engineering can induce complex collective behaviour and interesting synthetic gauge-field in quantum systems through temporal modulation of system parameters by periodic drives. Using a Floquet drive on frequencies of the magnon modes, we realize a chiral state-transfer in a cavity-magnonic system. The time-reversal symmetry is broken in such a promising platform for coherent information processing. In particular, the photon mode is adiabatically eliminated in the large-detuning regime and the magnon modes under conditional longitudinal drives can be indirectly coupled to each other with a phase-modulated interaction. The effective Hamiltonian is then used to generate chiral currents in a circular loop, whose dynamics is evaluated to measure the symmetry of the system Hamiltonian. Beyond the dynamics limited in the manifold with a fixed number of excitations, our protocol applies to the continuous-variable systems with arbitrary states. Also it is found to be robust against the systematic errors in the photon-magnon coupling strength and Kerr nonlinearity.
\end{abstract}

\date{\today}

\maketitle

\section{Introduction}\label{Introduction}

With unique properties observed in experiments such as large tunability and long coherent-time, the magnon can be used as information carrier in a broad variety of hybrid systems~\cite{cavitymagnonics,magnon,magnon2,magnonstate}. Found to be directly or effectively coupled to microwave photon~\cite{cavityyig1,cavityyig2,yigcavity,yigcavity2,kerrmagnon,gatemagnon}, superconducting qubit~\cite{magnonqubit2,magnonqubit,magnonqubit3,magnonqubit4}, spin-emitter~\cite{magnonspin,magnonspin2}, and phonon~\cite{magnoncavity,mppentang,intermagnon,quantumnetwork,magnonphonononchip,dynamicalmagnon,kerrmagnomech}, the magnon modes provide novel applications in quantum computing~\cite{quantumcomputing}, quantum communication~\cite{quantumcommunication}, and quantum sensing~\cite{quantumsense}. Fundamentally, the coupling between magnon and other types of mode motivates a number of insights, e.g., the indirect two-body flip-flop interaction of magnons for Rabi oscillation, which is symmetrical to time and parity. Moreover, Floquet engineering has been applied to a cavity electromagnonic system~\cite{Floquetmagnon}. It could support more flexible and interesting applications based on the complex-valued hopping, such as generating chiral transfers of the quasiparticles of spin waves as a highly promising bosonic platform.

Chirality has played an important role in the photon and phonon systems~\cite{chiral,chiral2}. Optical lattice could be used to simulate the chirality in magnetic materials~\cite{fraction}. In the platform of superconducting qubits~\cite{chiralspin2}, spin chirality arises from the three-body interactions. The chirality operator of three spins is $\hat{O}=\vec{\sigma}_1\cdot(\vec{\sigma}_2\times\vec{\sigma}_3)$, where $\vec{\sigma}_j\equiv(\sigma^x_j, \sigma^y_j, \sigma^z_j)$ is the Pauli vector for the $j$th spin particle. It is straightforward to verify that $\hat{O}$ breaks both the time reversal symmetry $\mathcal{T}$ (replacing $\vec{\sigma}_j$ by $-\vec{\sigma}_j$) and the parity symmetry $\mathcal{P}$ (exchanging $\vec{\sigma}_j$ with $\vec{\sigma}_k$) but conserves the $\mathcal{PT}$ symmetry~\cite{chiralspin2}. The dynamics of the spins driven by $\hat{O}$ features a chiral evolution, i.e., $|s_1s_2s_3\rangle\to$ $|s_2s_3s_1\rangle\to$ $|s_3s_1s_2\rangle$, where $s_j=0$ (spin down) or $1$ (spin up)~\cite{topolight}. In parallel to the protocol in the superconducting qubits, we propose to generate a chiral transfer of an arbitrary bosonic state in the Floquet cavity-magnonic system~\cite{Floquetmagnon}. Here the system-Hamiltonian symmetry is described by a nonvanishing chiral current operator.

According to the Floquet theory~\cite{Floquet,Floquet1,Floquettheory,Floquet3}, the propagator $U(t)$ induced by a time-dependent Hamiltonian $H(t)=H(t+T)$ with a period $T$ can be expressed by $U(t)=U_F(t)\exp(-iH't)$, where $U_F(t)$ is also periodic with $T$. As a time-independent Hermitian operator, $H'=U^{\dagger}_F(t)H(t)U_F(t)-iU^{\dagger}_F(t)\dot{U}_F(t)$ can be understood as the system Hamiltonian in the frame defined by $U_F(t)$. Floquet engineering by fast periodic modulation over the characteristic frequencies of a quantum system is a major control approach to the long-time dynamics of the system~\cite{Floquet1,James,Floquetnoon,Floquet2}, where the effective Hamiltonian can be synthesized through designing proper drives. Floquet engineering has been applied to quantum switch~\cite{switch}, chiral ground state current~\cite{chiralcurrent}, quantum simulation~\cite{Floquetsimu}, and perfect state transfer~\cite{chiralspin,chiralspin2,chiralspin3} in superconducting qubits. Also it found a very wide application in quantum gas systems~\cite{floquetgas,floquetgas2} to reveal many interesting physics.

In this work, we apply the Floquet engineering to a hybrid photon-magnon system in which three single-crystal yttrium iron garnet (YIG) spheres are placed inside a microwave cavity. The uniform bias magnetic field excites the Kittel mode in the YIG spheres and establishes a strong photon-magnon coupling. In the large detuning regime, the common photon mode can be adiabatically eliminated with the standard perturbation theory~\cite{secondorder,ae,ae1} or the high-order Fermi golden rule~\cite{fermigolden}. By periodically modulating over the three magnon modes with well-controlled intensities, frequencies, and phases, we can obtain an effective time-reversal symmetry broken Hamiltonian~\cite{timereversal}, that ensures chiral magnon currents~\cite{chiralcurrent} of arbitrary states with a state-independent period. In contrast to the protocols based on the superconducting qubits~\cite{chiralcurrent,chiralspin,chiralspin2}, our protocol is not constrained in a manifold of a fixed number of excitations and is capable to transfer versatile states such as Fock state, coherent state, Schr\"odinger cat state, and even two-body entangled states in a chiral way. The magnon Fock state can be generated by the magnon blockade~\cite{Fock1,Fock2}. A magnonic cat state was proposed in Ref.~\cite{catstate}. And more magnonic-state preparation protocols have been collected in a recent review~\cite{magnonstate}.

The rest part of this work is structured as follows. In Sec.~\ref{secmodel}, we introduce the quantum model for a Floquet cavity-magnonic system and then derive an effective Hamiltonian to generate a perfect chiral state transfer amongst magnons. Clockwise and anticlockwise transfers of various states and chiral currents to measure the symmetry of the system Hamiltonian are presented in Sec.~\ref{statetransfer} and Sec.~\ref{seccurrent}, respectively. State-transfer fidelity under the magnon damping and the systematic errors arise from the photon-magnon coupling strength and the Kerr effects are discussed in Sec.~\ref{damping} and Sec.~\ref{syserror}, respectively. The whole work is summarized in Sec.~\ref{conclu}.

\section{Model}\label{secmodel}

Consider a hybrid quantum model consists of a microwave cavity coupled to $N$ YIG spheres in their Kittle modes~\cite{Floquetmagnon,kerrmagnon,kerrmagnon2}. The model Hamiltonian reads
\begin{equation}\label{Ham}
H=\omega_aa^{\dagger}a+\omega_m\sum_{k=1}^Nm^{\dagger}_km_k+g_{am}\sum_{k=1}^N \left(am^{\dagger}_k+a^{\dagger}m_k\right),
\end{equation}
where $a$ $(a^{\dagger})$ and $m_k$ $(m^{\dagger}_k)$ are the annihilation (creation) operators of the photon and the $k$th magnon modes, respectively. $\omega_a$ and $\omega_m$ are their respective transition frequencies. $g_{am}$ is the single-excitation coupling strength between photon and magnon modes, which is much larger than their decay rates and much smaller than their detuning $|\omega_a-\omega_m|$ in the dispersive regime. Counter-rotating interaction between photon and magnons is omitted under the rotating-wave approximation (RWA), as justified in Appendix~\ref{appa}.

When one focuses on the state exchange within any pair of magnon modes, it is instructive to show that the photon mode could be adiabatically eliminated by the second-order perturbation~\cite{secondorder}. The subsystem Hamiltonian associated with the $k$th and $j$th magnon modes and the photon mode can be written as $H^{kj}_0+H^{kj}_I$, where
\begin{equation}\label{Hsub}
\begin{aligned}
H^{kj}_0&=\omega_a a^{\dagger} a+\omega_m\left(m_k^{\dagger}m_k+m_j^{\dagger}m_j\right), \\
H^{kj}_I&=g_{am}\left[a\left(m^{\dagger}_k+m^{\dagger}_j\right)+a^{\dagger}\left(m_k+m_j\right)\right]
\end{aligned}
\end{equation}
indicating the unperturbed and perturbation Hamiltonians, respectively. To the second order in the perturbation Hamiltonian, the effective coupling strength $g_{pq}$ between any pair of eigenstates $|p\rangle$ and $|q\rangle$ of $H^{kj}_0$ in Eq.~(\ref{Hsub}) is given by
\begin{equation}\label{geffper}
g_{pq}=\sum_{w\neq p,q}\frac{\langle q|H^{kj}_I|w\rangle\langle w|H^{kj}_I|p\rangle}{E_p-E_w},
\end{equation}
where $E_w$ is the eigenenergy of an intermediate eigenstate $|w\rangle$, namely, $H^{kj}_0|w\rangle=E_w|w\rangle$. Note that the large detuning condition $\omega_m\neq\omega_a$ lifts the degeneracy of $H^{kj}_0$. Due to the excitation conservation in both Eqs.~(\ref{Ham}) and (\ref{Hsub}), $g_{pq}$ can be probed as the transition rate between $|p\rangle=|n_kn_an_j\rangle\equiv|n_k\rangle|n_a\rangle|n_j\rangle$ and $|q\rangle=|(n_k-1)n_a(n_j+1)\rangle$ in a subspace with a fixed number of excitations $M=n_k+n_a+n_j$, where the subscript indicates the mode and $n_k$, $n_a$, and $n_j$ are arbitrary nonnegative integers. By Eqs.~(\ref{Hsub}) and (\ref{geffper}), the leading contribution for $|n_kn_an_j\rangle\to|(n_k-1)n_a(n_j+1)\rangle$ comes from two physical paths, i.e., $|n_kn_an_j\rangle\to|(n_k-1)(n_a+1)n_j\rangle\to|(n_k-1)n_a(n_j+1)\rangle$ and $|n_kn_an_j\rangle\to|n_k(n_a-1)(n_j+1)\rangle\to|(n_k-1)n_a(n_j+1)\rangle$. Then in this subspace we have
\begin{equation}\label{tildeg}
g_{n_kn_j}=\sqrt{n_k(n_j+1)}\frac{g^2_{am}}{\omega_m-\omega_a}=g\sqrt{n_k(n_j+1)}.
\end{equation}
Note $m_k|n_k\rangle=\sqrt{n_k}|n_k-1\rangle$ and $m_j^{\dagger}|n_j\rangle=\sqrt{n_j+1}|n_j+1\rangle$. The effective Hamiltonian for this pair of magnon modes in their full space~\cite{intermagnon} can thus be written as
\begin{equation}\label{tildeH}
H_{kj}=g\left(m_km^{\dagger}_j+m^{\dagger}_km_j\right), \quad g\equiv\frac{g^2_{am}}{\omega_m-\omega_a}.
\end{equation}

The preceding derivation as well as the construction of the effective Hamiltonian applies to any indirect magnon-magnon interaction mediated by the common photon mode. In the dispersive regime, the Hamiltonian for the whole model in Eq.~(\ref{Ham}) can therefore be written as
\begin{equation}\label{Heffsecond}
H=g\sum_{k<j}^N\left(m_km^{\dagger}_j+m^{\dagger}_km_j\right).
\end{equation}
We apply the periodic signals by Floquet driving~\cite{Floquetmagnon,chiralspin2} featured with the intensity $\Delta$, the frequency $\omega$, and the local phases $\phi_k$, to the model Hamiltonian and obtain
\begin{equation}\label{Hamiltonian}
\begin{aligned}
H(t)&=\Delta\sum_{k=1}^N\cos(\omega t+\phi_k)m^{\dagger}_km_k\\
&+g\sum_{k<j}^N\left(m_km^{\dagger}_j+m^{\dagger}_km_j\right).
\end{aligned}
\end{equation}

In the rotating frame with respect to
\begin{equation}\label{rotationU}
\begin{aligned}
U_0(t)&=\exp\left[i\int^t_0ds\Delta\sum_{k=1}^N\cos(\omega s+\phi_k)m^{\dagger}_km_k\right]\\
&=\exp\left\{i\sum^N_{k=1}\frac{\Delta}{\omega}\left[\sin(\omega t+\phi_k)-\sin\phi_k\right]m^{\dagger}_km_k\right\},
\end{aligned}
\end{equation}
we have
\begin{equation}\label{interHam}
\begin{aligned}
H_I(t)&=U_0(t)H(t)U^{\dagger}_0(t)-iU_0(t)\dot{U}^{\dagger}_0(t)\\
=&\sum_{k<j}^Ng_{kj}\exp\left[if_{kj}\sin(\omega t+\alpha_{kj})\right]m_k^{\dagger}m_j+{\rm H.c.},
\end{aligned}
\end{equation}
where
\begin{equation*}
\begin{aligned}
& g_{kj}=ge^{-i\beta_{kj}}, \quad \beta_{kj}\equiv\frac{\Delta}{\omega}(\sin\phi_k-\sin\phi_j), \\
& f_{kj}=\frac{2\Delta}{\omega}\sin\left(\frac{\phi_j-\phi_k}{2}\right), \\
& \alpha_{kj}\equiv\tan^{-1}\left(\frac{\sin\phi_k-\sin\phi_j}{\cos\phi_k-\cos\phi_j}\right).
\end{aligned}
\end{equation*}

According to the Jacobi-Anger expansion $e^{ix\sin y}=\sum_{n=-\infty}^{n=+\infty}J_n(x)e^{iny}$, where $J_n$ is the $n$th Bessel function of the first kind, one can obtain
\begin{equation}\label{interHamn}
H_I(t)=H_0+\sum_{n=1}^{\infty}\left(H_ne^{in\omega t}+H_{-n}e^{-in\omega t}\right),
\end{equation}
with
\begin{equation}\label{H0Hn}
\begin{aligned}
H_0&=\sum_{k<j}^N J_0(f_{kj})\left(g_{kj}m^{\dagger}_km_j+g^*_{kj}m_km_j^{\dagger}\right), \\
H_n&=\sum_{k<j}^N J_n(f_{kj})e^{in\alpha_{kj}}\left[g_{kj}m^{\dagger}_km_j+(-1)^ng_{kj}^*m_km_j^{\dagger}\right],
\end{aligned}
\end{equation}
and $H_{-n}=H_n^*$. Therefore up to the order of $\mathcal{O}(1/\omega)$, the Floquet-driving Hamiltonian~(\ref{Hamiltonian}) can be written as
\begin{equation}\label{effHam0}
H_{\rm eff}=H_0+H_{\rm eff}^{(2)}=H_0+\sum^{\infty}_{n=1}\frac{1}{n\omega}\left[H_n, H_{-n}\right].
\end{equation}
We have several remarks for $H_{\rm eff}$. (a) The zeroth-order coupling-strength between the $k$th and $j$th magnons is $g_{kj}J_0(f_{kj})$, which can be tuned by the ratio of the driving intensity and frequency $\Delta/\omega$ and the phase difference between the local driving signals $\phi_k-\phi_j$. (b) When $N=2$, the leading-order correction by the commutator $[H_n, H_{-n}]$ vanishes and then the dynamics of magnons becomes a Rabi oscillation with a frequency fully determined by $f_{kj}$. (c) When $N>2$, the leading-order correction can be written as
\begin{equation}
H_{\rm eff}^{(2)}=\sum_{k<j}\left(g^{(2)}_{kj}e^{i\alpha_{kj}}m_k^{\dagger}m_j+h.c.\right),
\end{equation}
where the second-order coupling strength is
\begin{equation}
g^{(2)}_{kj}\sim\frac{g^2}{\omega}\sum_{k_1\neq k,j}\left[\sum^{\infty}_{n=1}\frac{J_n(f_{kk_1})J_n(f_{k_1j})}{n}\right],
\end{equation}
with $k_1$ indicating the other magnon modes rather than $k$ and $j$ in the whole system.

The third and even higher orders of terms can be analyzed with the James' effective-Hamiltonian method~\cite{James}. The nonvanishing third-order terms in the effective Hamiltonian can be obtained by
\begin{equation}\label{effHam3}
\begin{aligned}
H^{(3)}_{kj}&=\sum_{l=m+n}\frac{1}{(n+m)\omega^2}\left[H_n, \left[H_m, H_{-l}\right]\right] \\
&=\sum_{k<j}\left(g^{(3)}_{kj}e^{i\alpha_{kj}}m_k^{\dagger}m_j+H.c.\right),
\end{aligned}
\end{equation}
where the third-order coupling strength $g^{(3)}_{kj}$ is about
\begin{equation}
\sim\frac{g^3}{\omega^2}\sum_{k_1\neq k, k_2\neq j}\left[\sum_{l=m+n}\frac{J_n(f_{kk_1})J_m(f_{k_1k_2})J_l(f_{k_2j})}{n+m}\right].
\end{equation}
Therefore the third and even higher orders of corrections can be safely omitted in the dispersive regime $g\ll\omega$.

Chiral state-transfer (as described in Fig.~\ref{model}) could manifest when $N=3$ by setting $\phi_j=2\pi j/3$, $j=1,2,3$, which are uniformly distributed in the range of $[0, 2\pi]$. In this case, the magnitude of the ratio factor $f_{kj}$ becomes independent of the magnon pair, i.e., $f_{12}=f_{13}=f_{23}=\sqrt{3}\Delta/\omega\equiv f$. The relevant quantum phases are found to be $\beta_{12}=f$, $\beta_{23}=-f/2$, $\beta_{13}=f/2$; and $\alpha_{12}=\pi/2$, $\alpha_{13}=5\pi/6$, $\alpha_{23}=7\pi/6$, i.e., $\alpha_{kj}=(k+j)\pi/3-\pi/2$. Then using $H_n$ in Eq.~(\ref{H0Hn}), the second-order term in the effective Hamiltonian~(\ref{effHam0}) turns out to be
\begin{equation}\label{Heff}
H^{(2)}_{\rm eff}=-ig_{\rm eff}\sum_{j=k+1}\left(e^{-i\beta_{kj}}m_k^{\dagger}m_j-e^{i\beta_{kj}}m_km^{\dagger}_j\right).
\end{equation}
Note if $k=3$, then $j=1$, and
\begin{equation}
g_{\rm eff}=g_{\rm eff}(f)=\frac{2g^2}{\omega}\sum_{n=1}^{\infty}\frac{J^2_n(f)}{n}\sin\left(\frac{n\pi}{3}\right).
\end{equation}
Consequently, the effective Hamiltonian in Eq.~(\ref{effHam0}) can be expressed by an coefficient matrix,
\begin{equation}\label{Heff0matrix}
H_{\rm eff}=[m^{\dagger}_1,m^{\dagger}_2,m^{\dagger}_3]\begin{bmatrix}
0 & G^*e^{-if} & Ge^{-if/2}\\
Ge^{if} & 0 &G^*e^{if/2}\\
G^*e^{if/2} & Ge^{-if/2} & 0
\end{bmatrix}\begin{bmatrix}
m_1\\m_2\\m_3\end{bmatrix},
\end{equation}
where $G=gJ_0(f)+ig_{\rm eff}$ is the effective coupling strength. Using $\tan\phi\equiv g_{\rm eff}/[gJ_0(f)]$, the Hamiltonian has a more compact form,
\begin{equation}\label{Heff0phi}
H_{\rm eff}=|G|[m^{\dagger}_1,m^{\dagger}_2,m^{\dagger}_3]\begin{bmatrix}
0 & e^{i\phi_{12}} & e^{-i\phi_{31}}\\
e^{-i\phi_{12}} & 0 &e^{i\phi_{23}}\\
e^{i\phi_{31}} & e^{-i\phi_{23}} & 0
\end{bmatrix}\begin{bmatrix}
m_1\\m_2\\m_3\end{bmatrix},
\end{equation}
where $\phi_{12}=-\phi-f$, $\phi_{23}=-\phi+f/2$, and $\phi_{31}=-\phi+f/2$. The eigenvalues of the coefficient matrix satisfy $E^3-3E-2\cos\Phi=0$, where $\Phi\equiv\phi_{12}+\phi_{23}+\phi_{31}=-3\phi$ is defined as the closed-loop phase up to $2n\pi$ with $n$ an integer. The system dynamics is then found to be relevant to the driving intensity and frequency rather than solely determined by the ratio factor $f$. $\Phi$ is thus regarded as the synthetic magnetic flux. It is gauge-invariant by noting that in our model the three YIGs form a closed loop and the accumulated phase of the state-current (either chiral or not) has to be single-valued when going around this loop. $\Phi=0$ or $\pi$ corresponds to the absence of the Floquet engineering. Then the evolution of the magnons is symmetrical when one of the magnons (say mode-$1$) is prepared at the target state and the rest two are in the same states. The quantum state propagates from mode-$1$ to mode-$2$ and mode-$3$ simultaneously and then back to mode-$1$. This pattern repeats itself with no indication of any preferred circulation direction.

\section{Numerical simulation of Chiral currents}\label{result}

\subsection{Chiral state transfer}\label{statetransfer}

\begin{figure}[htbp]
\centering
\includegraphics[width=0.48\textwidth]{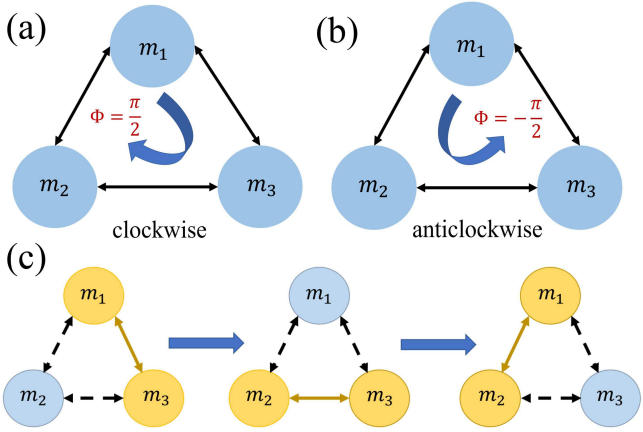}
\caption{(a) and (b): Diagrams of the chiral state transfers in the magnons along the clockwise and anticlockwise circles, respectively. (c): Diagram of the chiral entanglement transfer along the clockwise direction, where two yellow nodes connected by a solid line represent the two entangled magnons and the blue node represents a separable magnon.}\label{model}
\end{figure}

With a Hamiltonian similar to Eq.~(\ref{Heff0phi}), that is obtained by the tunneling-term modulation~\cite{chiralcurrent} rather than the longitudinal Floquet driving, the time-reversal symmetry broken was distinguished when the loop phase $\Phi=\pm\pi/2$. This idea can be generalized to realize a chiral state-transfer in our continuous-variable system for arbitrary target states.

Due to Eq.~(\ref{Heff0matrix}), $\Phi=\pi/2$ yields $G=ig_{\rm eff}$ under the condition of $J_0(f)=0$. It renders that $f=\sqrt{3}\Delta/\omega=2.4048$. Consequently, $J_1(f)=0.5192$, $J_2(f)=0.4318$, $J_3(f)=0.1990$, $J_4(f)=0.0647$, $J_5(f)=0.0164$, $J_6(f)=0.0034$. The Hamiltonian becomes
\begin{equation}\label{Heffmatrix}
H_{\rm eff}=ig_{\rm eff}[m^{\dagger}_1,m^{\dagger}_2,m^{\dagger}_3]\begin{bmatrix}
0 & -e^{-if} & e^{-if/2}\\
e^{if} & 0 &-e^{if/2}\\
-e^{if/2} &e^{-if/2} & 0
\end{bmatrix}\begin{bmatrix}
m_1\\m_2\\m_3\end{bmatrix}.
\end{equation}
Note when $N\geq4$, $f_{kj}$'s are not the same in magnitude. Consequently, the zeroth-order time-reversal symmetry Hamiltonian $H_0$ in Eq.~(\ref{H0Hn}) remains in any case. It is then hard to observe a perfect chiral state-transfer. The canonical transformation in the Heisenberg picture induced by Eq.~(\ref{Heffmatrix}) can be written as
\begin{equation}\label{mUm}
\begin{bmatrix}
m_1(t) \\ m_2(t) \\ m_3(t)
\end{bmatrix}=T(t)\begin{bmatrix}
m_1(0) \\ m_2(0) \\  m_3(0)
\end{bmatrix}
\end{equation}
with
\begin{equation}\label{HeisenbergUt}
T(t)=\frac{1}{3}\begin{bmatrix}
x(t) & e^{-if}y(t) & e^{-\frac{if}{2}}z(t)\\
e^{if}z(t) & x(t) & e^{\frac{if}{2}}y(t)\\
e^{\frac{if}{2}}y(t) & e^{-\frac{if}{2}}z(t) & x(t)
\end{bmatrix},
\end{equation}
where
\begin{equation}\label{xyz}
\begin{aligned}
x(t)&=1+2\cos\left(\sqrt{3}g_{\rm eff}t\right), \\
y(t)&=1-2\cos\left(\sqrt{3}g_{\rm eff}t+\frac{\pi}{3}\right),\\
z(t)&=1-2\cos\left(\sqrt{3}g_{\rm eff}t-\frac{\pi}{3}\right).
\end{aligned}
\end{equation}
With respect to the time-evolved operator, it is interesting to find that $m_3(t)=m_1(0)$ and $m_2(t)=m_3(0)$ when $t=(2\pi/3+2n\pi)/(\sqrt{3}g_{\rm eff})$; $m_2(t)=m_1(0)$ and $m_3(t)=m_2(0)$ when $t=(4\pi/3+2n\pi)/(\sqrt{3}g_{\rm eff})$; and $m_j(t)=m_j(0)$ when $t=2n\pi/(\sqrt{3}g_{\rm eff})$ with $n$ an integer. This transfer is exactly described by the clockwise rotation $m_1\to m_3\to m_2\to m_1$ in Fig.~\ref{model}(a), corresponding to a chiral evolution in the Schr\"odinger picture, i.e., $|\varphi_1\varphi_2\varphi_3\rangle\to$ $|\varphi_2\varphi_3\varphi_1\rangle\to$ $|\varphi_3\varphi_1\varphi_2\rangle$, where $\varphi_j$ is arbitrary for the $j$th mode.

The state-transfer fidelity in the Schr\"odinger picture can be measured by the time-dependent state population $P_j(t)$, $j=1,2,3$, which can be defined by
\begin{equation}\label{population}
P_j(t)=\sum_{C_n\neq 0}\left|\left\langle\varphi(t)\Big|\frac{(m^{\dagger}_j)^n}{\sqrt{n!}}\Big|000\right\rangle\right|^2.
\end{equation}
It describes the overlap between the target state and the $j$th mode's state without regarding the influence from local dynamical phases~\cite{population}. Here the initial state of the magnons is assumed to be $|\varphi(0)\rangle=\sum_nC_n|n00\rangle\equiv\sum_nC_n|n\rangle_1|0\rangle_2|0\rangle_3$ with the normalized coefficients $C_n$, i.e., the first magnon is prepared as an arbitrary target state and the other two are in their ground states. The dynamics of the magnons is analytically determined by Eq.~(\ref{mUm}) or numerically calculated by the Floquet-driving Hamiltonian in Eq.~(\ref{Hamiltonian}).

For $|\varphi(0)\rangle=|100\rangle$, we have
\begin{equation}\label{timestate1}
|\varphi(t)\rangle=\frac{1}{3}\left[x(t)|100\rangle+e^{-if}z(t)|010\rangle+e^{\frac{-if}{2}}y(t)|001\rangle\right].
\end{equation}
While for an arbitrary superposed state
\begin{equation}\label{state0}
|\varphi(0)\rangle=\sum_nC_n|n00\rangle=\sum_n\frac{C_n}{\sqrt{n!}}\left[m^{\dagger}_1(0)\right]^n|000\rangle,
\end{equation}
the time-evolved state can be written as
\begin{equation}\label{statet}
|\varphi(t)\rangle=\sum_n\frac{C_n}{\sqrt{n!}}\left[\sum_jm^{\dagger}_1(0)T_{j1}^{\dagger}(t)\right]^n|000\rangle.
\end{equation}
Then due to Eqs.~(\ref{mUm}) and (\ref{HeisenbergUt}), we have $\varphi(t_3)=\sum_nC_n|00n\rangle$ at the desired time $t_3=2\pi/(3\sqrt{3}g_{\rm eff})$ and $\varphi(t_2)=\sum_nC_n|0n0\rangle$ at $t_2=4\pi/(3\sqrt{3}g_{\rm eff})$, up to certain local phases. By virtue of Eq.~(\ref{population}), we have $P_1=\sum_nx^{2n}(t)|C_n|^2/9^n$, $P_2=\sum_nz^{2n}(t)|C_n|^2/9^n$, and $P_3=\sum_ny^{2n}(t)|C_n|^2/9^n$. During the state transfer of $m_1\to m_3$, the population of $m_2$ achieve the side peak value at $t=(\pi/3+2n\pi)/(\sqrt{3}g_{\rm eff})$, which is undesired for $m_1\to m_3$ yet cannot be under control. In particular, $P_2^{\rm max}=\sum_n|C_n|^2/9^n$ at those moments. Similarly, when $t=(\pi+2n\pi)/(\sqrt{3}g_{\rm eff})$ and $t=(5\pi/3+2n\pi)/(\sqrt{3}g_{\rm eff})$, $P_1$ and $P_3$ are expected to achieve their side peak values in the middle of $m_3\to m_2$ and $m_2\to m_1$, respectively. In addition, one can even transfer a mixed state by $\rho(t)=\sum_jp_j|\varphi(t)\rangle_{jj}\langle\varphi(t)|$ based on Eq.~(\ref{statet}).

\begin{figure}[htbp]
\centering
\includegraphics[width=0.48\textwidth]{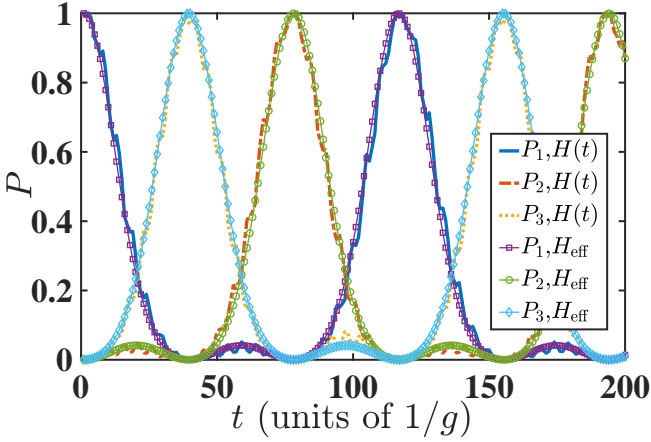}
\caption{Time evolutions of the state populations $P_j(t)$, $j=1,2,3$, for the three magnons by the numerical simulation with the Floquet-driving Hamiltonian~(\ref{Hamiltonian}) and the analytical evaluation with the canonical transformation~(\ref{mUm}). The initial state is $|\varphi(0)\rangle=1/\sqrt{3}\sum_{n=1}^3|n00\rangle$. $\omega=20g$ and $\phi_j=2\pi j/3$.}\label{anavsnum}
\end{figure}

Figure~\ref{anavsnum} is used to verify the time evolutions of the state populations $P_j(t)$ by the Floquet-driving Hamiltonian~(\ref{Hamiltonian}) under the chiral condition of $J_0(f)=0$ in comparison with those by the effective Hamiltonian~(\ref{Heffmatrix}) or the canonical transformation~(\ref{mUm}). It is found that the analytical results (lines with markers) do match with their numerical counterparts (lines without markers) for the normalized time evolution. The periodic dynamics is determined once $g$ is fixed.

\begin{figure}[htbp]
\centering
\includegraphics[width=0.23\textwidth]{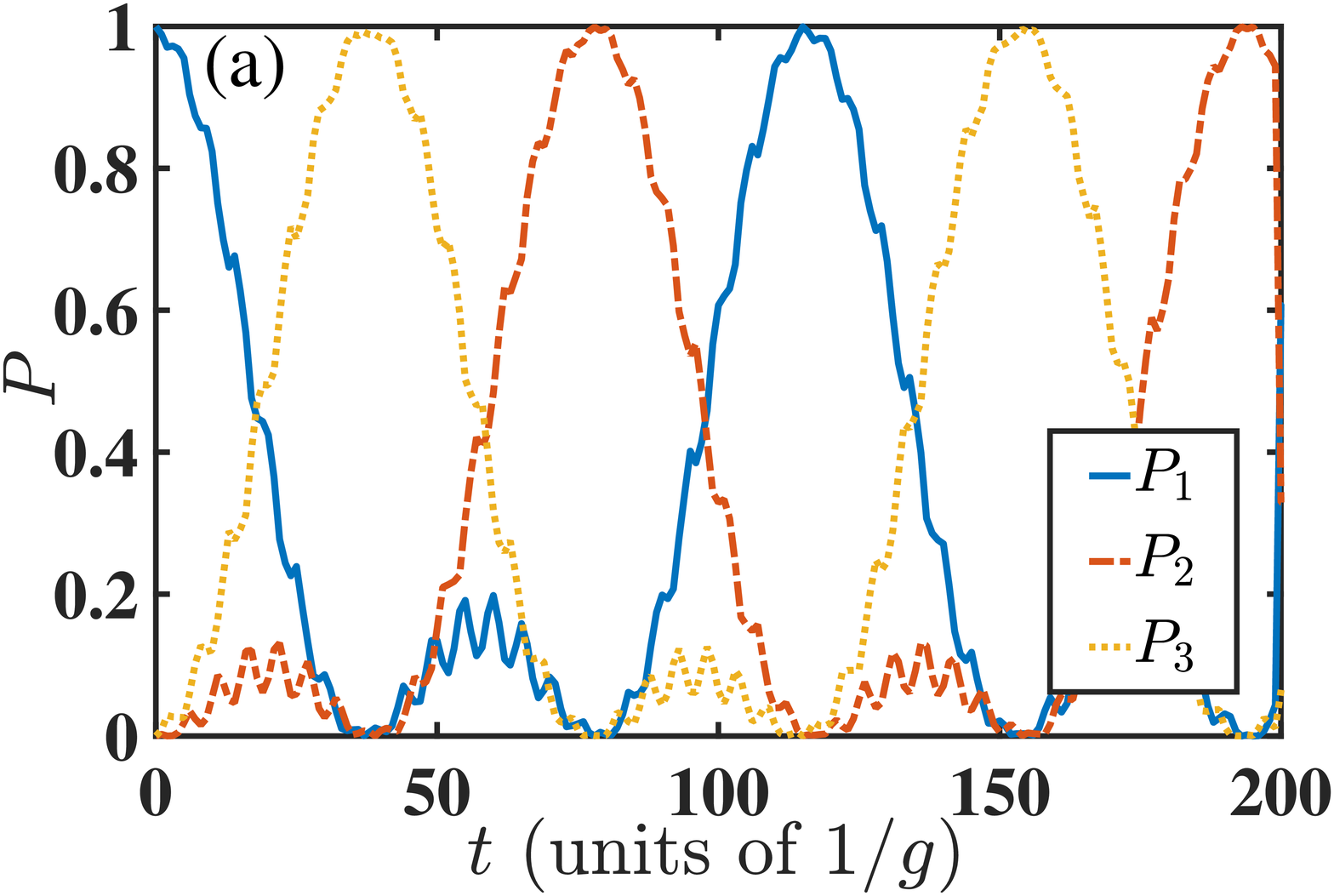}
\includegraphics[width=0.23\textwidth]{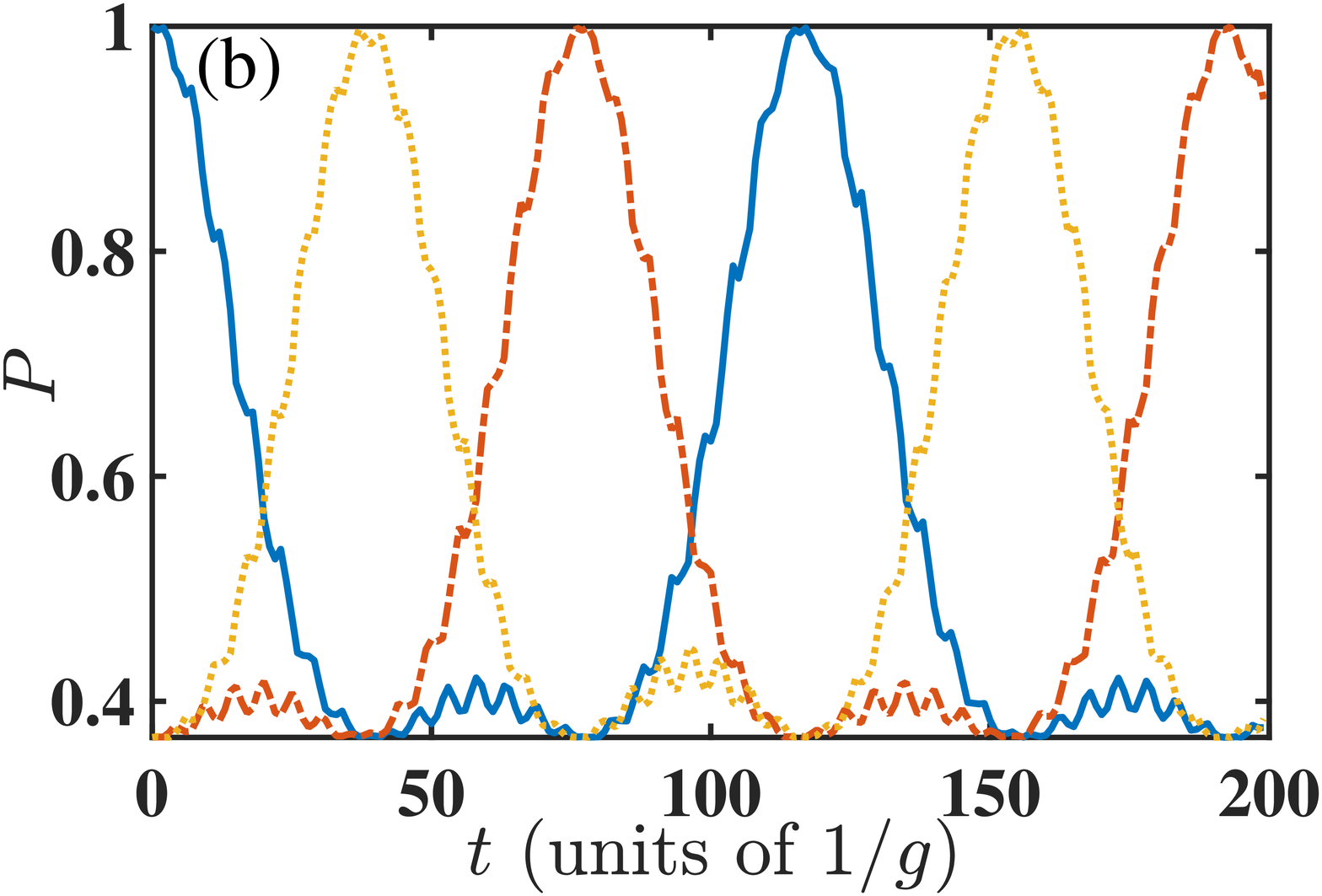}
\includegraphics[width=0.23\textwidth]{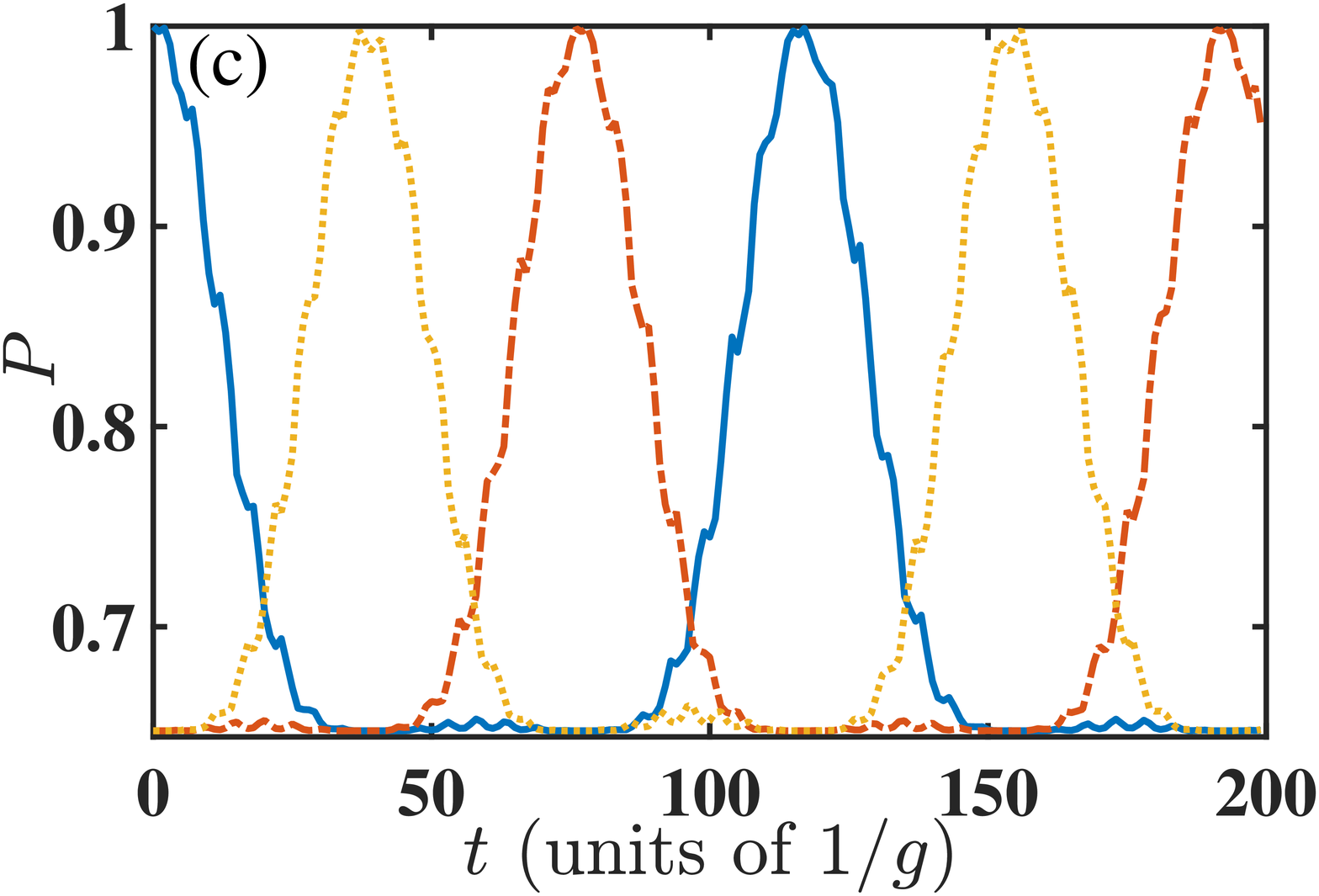}
\includegraphics[width=0.23\textwidth]{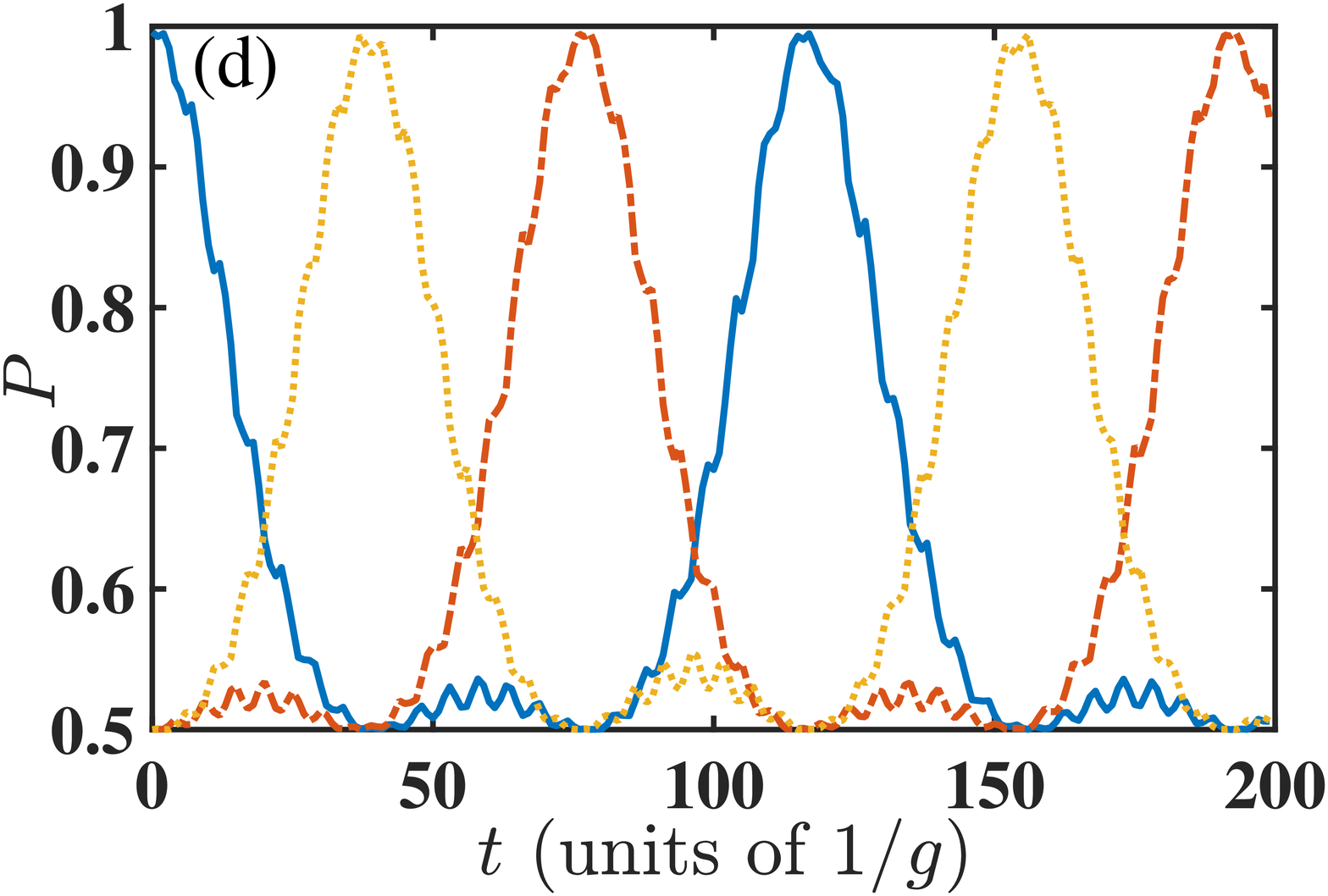}
\caption{Clockwise state-transfer of the three magnons described by $P_j$'s under the Floquet-driving Hamiltonian~(\ref{Hamiltonian}). The initial states for $m_1$ are (a) the Fock state $|n=1\rangle$, (b) the coherent state with $\beta=1$, (c) the cat state with $\zeta=1$, and (d) the thermal state with $\bar{n}=1$. $\omega=20g$ and $\phi_j=2\pi j/3$. }\label{result1}
\end{figure}

In Fig.~\ref{result1}, the chiral transfer protocol is applied to Fock state, Glauber coherent state $|\beta\rangle$, cat state $(|\zeta\rangle+|-\zeta\rangle)/\sqrt{2+2e^{-2|\zeta|^2}}$ with $|\zeta\rangle$ a coherent state, and thermal state $\rho=\sum_np_n|n\rangle\langle n|$ with the Fock-state occupation $p_n=(\bar{n})^n/(1+\bar{n})^{n+1}$ and the average excitation number $\bar{n}$. Periodically all the state populations approach unit in sequence indicating a perfect clockwise transfer amongst $m_1$, $m_2$, and $m_3$. It is interesting to find that the side peak for the Fock state is more significant than the other states that occupy the vacuum state. It is caused by the distinction between the side peak and initial values $P_2^{\rm max}-P_2(0)=\sum_n(1/9)^n|C_n|^2-|C_0|^2$. In Fig.~\ref{result1}(a), (b) and (c), the distinctions are about $0.11$, $0.05$ and $0.01$, respectively. Figures~\ref{anavsnum} and \ref{result1} demonstrate that the Floquet-driving Hamiltonian in Eq.~(\ref{Heffmatrix}) under the chiral conditions $J_0(f)=0$ and $\phi_j=2\pi j/3$ would ensure a high-fidelity chiral state-transfer for arbitrary target states. And the transfer period is irrespective to the chosen initial state.

\begin{figure}[htbp]
\centering
\includegraphics[width=0.23\textwidth]{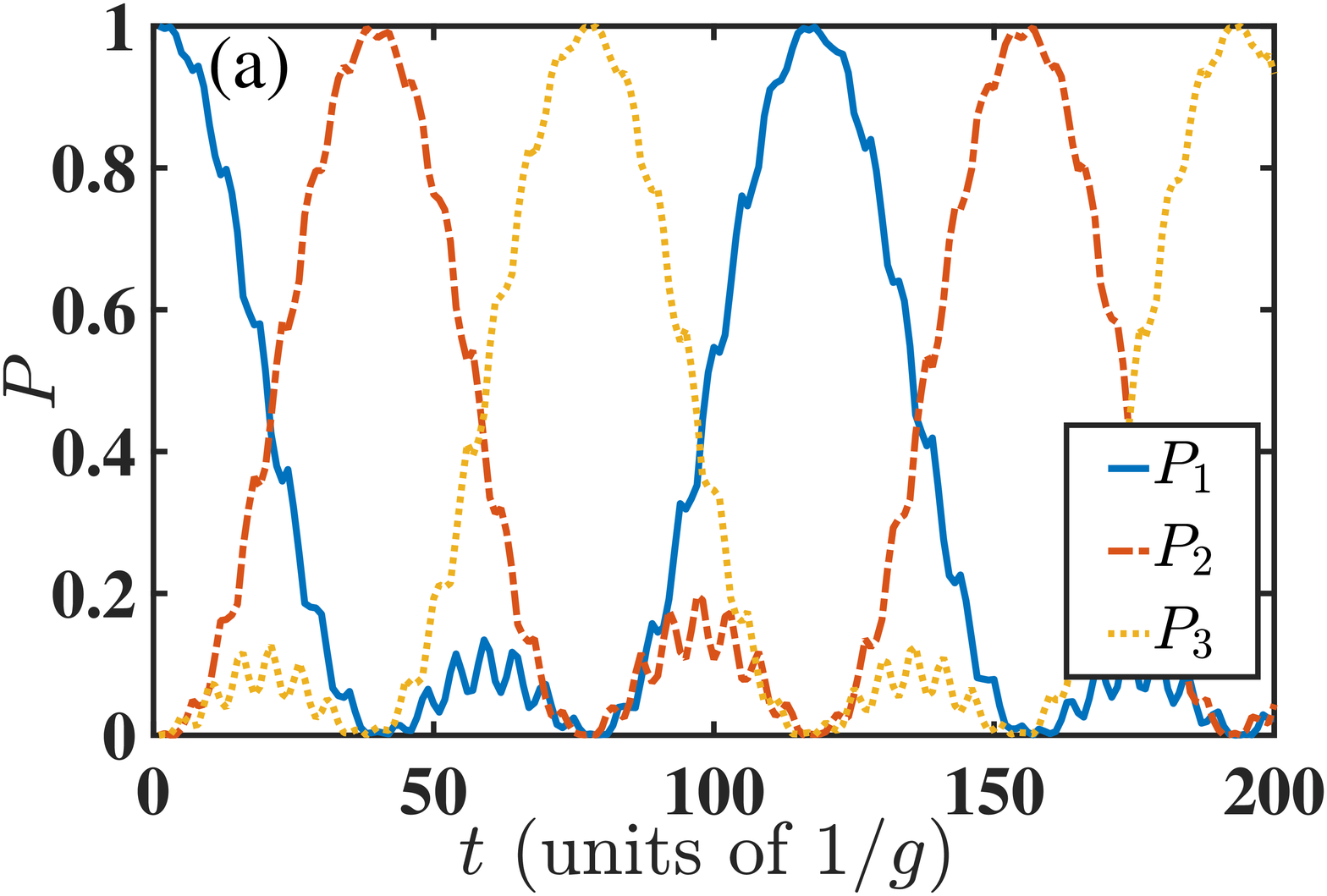}
\includegraphics[width=0.23\textwidth]{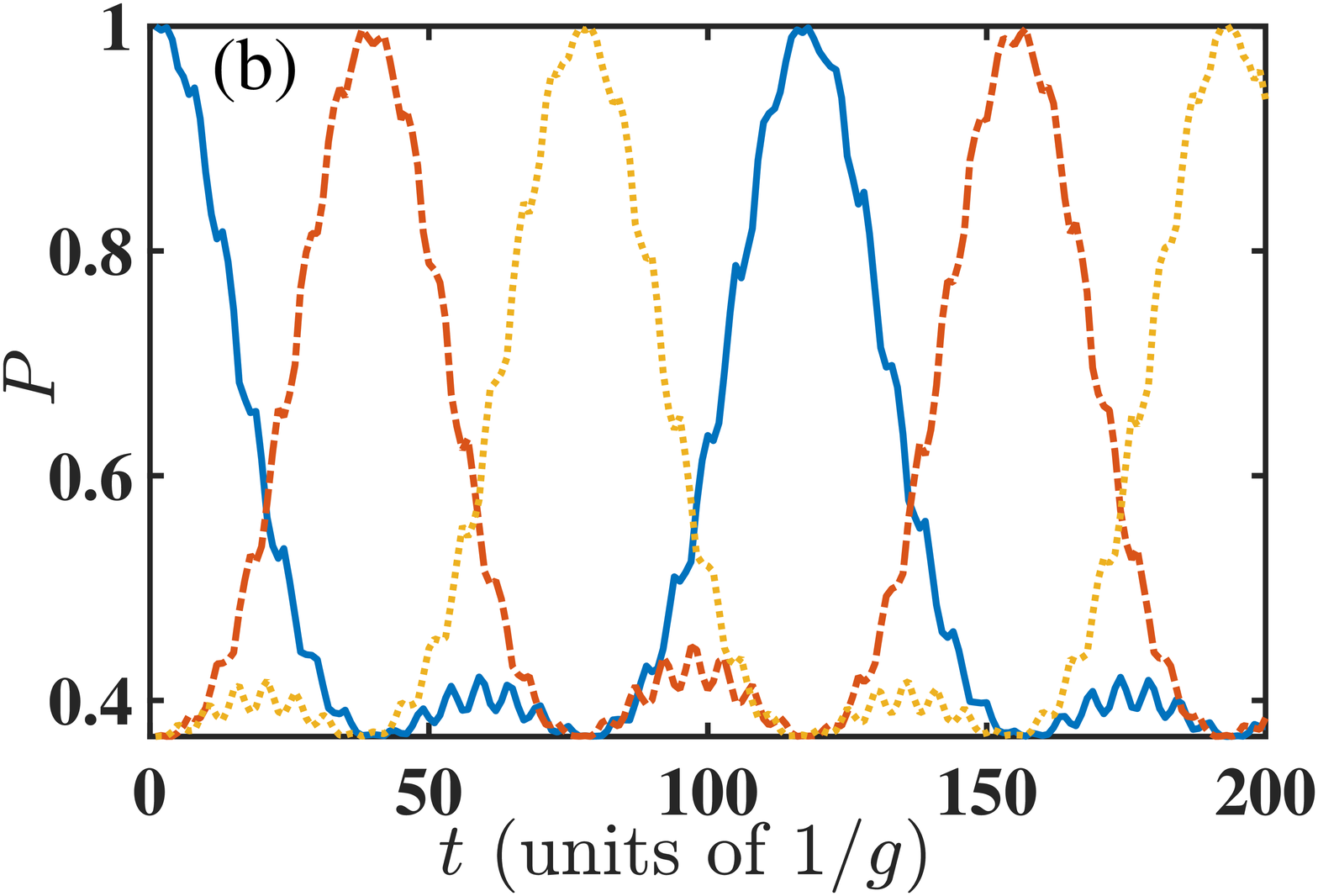}
\includegraphics[width=0.23\textwidth]{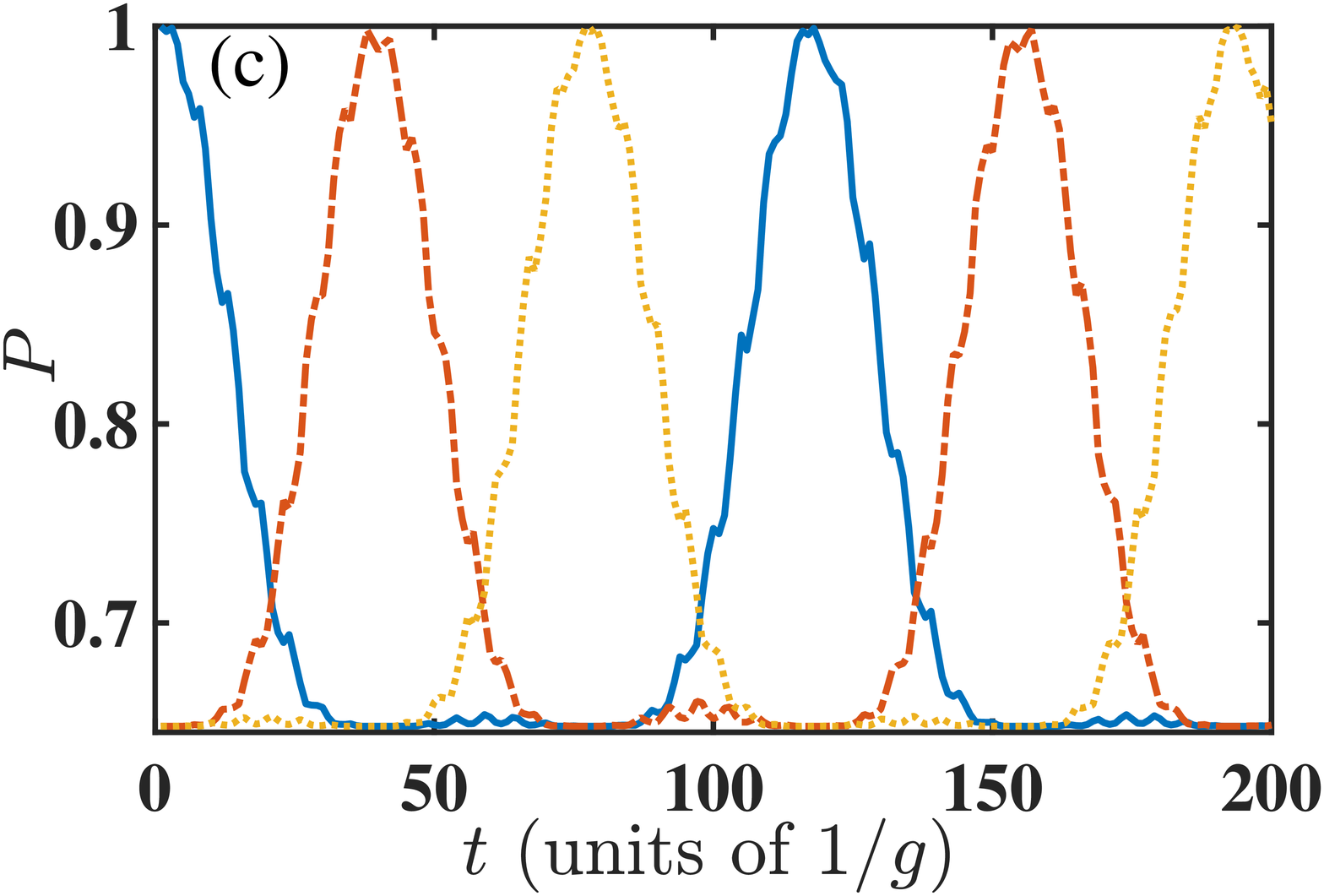}
\includegraphics[width=0.23\textwidth]{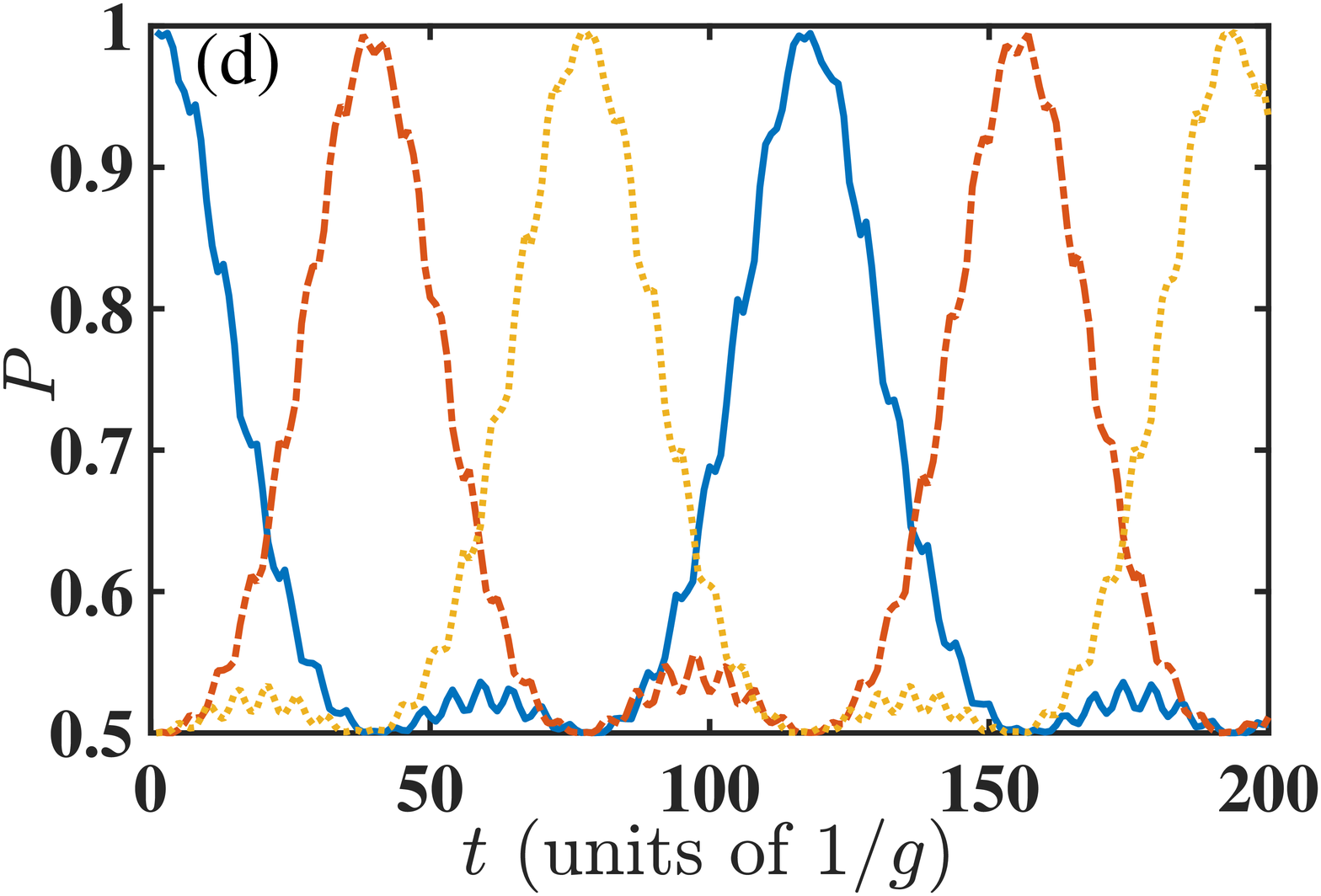}
\caption{Anticlockwise state-transfer of the magnons described by $P_j$'s under the Floquet-driving Hamiltonian~(\ref{Hamiltonian}). The initial states of the magnon-mode $m_1$ are (a) the Fock state $|n=1\rangle$, (b) the coherent state with $\beta=1$, (c) the cat state with $\zeta=1$, and (d) the thermal state with $\bar{n}=1$. $\omega=20g$, $\phi_1=2\pi/3$, $\phi_2=2\pi$, and $\phi_3=4\pi/3$.}\label{result2}
\end{figure}

By setting $\phi_1=2\pi/3$, $\phi_2=2\pi$, and $\phi_3=4\pi/3$, which is an alternative uniform distribution in the range of $[0, 2\pi]$, the coefficient matrix in Eq.~(\ref{Heff0matrix}) becomes
\begin{equation}\label{Heffmatrix1}
\begin{bmatrix}
0 & Ge^{-if/2} & G^*e^{-if}\\
G^*e^{if/2} & 0 &G^*e^{if/2}\\
Ge^{if} & Ge^{if/2} & 0
\end{bmatrix}
\end{equation}
where $G=gJ_0(f)-ig_{\rm eff}$. That induces the synthetic magnetic flux with $\Phi=-\pi/2$. It amounts to exchange the magnon mode-$2$ and $3$, giving rise to the anticlockwise rotation shown in Fig~\ref{model}(b). Then the effective Hamiltonian in Eq.~(\ref{Heffmatrix}) is modified to
\begin{equation}\label{Heffmatrix2}
H_{\rm eff}=ig_{\rm eff}[m^{\dagger}_1,m^{\dagger}_2,m^{\dagger}_3]\begin{bmatrix}
0 & e^{-if/2} & -e^{-if}\\
-e^{if/2} & 0 &-e^{-if/2}\\
e^{if} & e^{if/2} & 0
\end{bmatrix}\begin{bmatrix}
m_1\\m_2\\m_3\end{bmatrix}
\end{equation}
Through a canonical transformation similar to Eq.~(\ref{mUm}), one can easily analyse the anticlockwise state-transfer. For example, the time-dependent state $|\varphi(t)\rangle$ in Eq.~(\ref{timestate1}) for the single-excitation state becomes
\begin{equation}\label{timestate2}
|\varphi(t)\rangle=\frac{1}{3}\left[x(t)|100\rangle+e^{\frac{-if}{2}}y(t)|010\rangle+e^{-if}z(t)|001\rangle\right],
\end{equation}
where $x(t)$, $y(t)$, and $z(t)$ are given by Eq.~(\ref{xyz}). The numerical results for various states are provided in Fig.~\ref{result2}, exhibiting the opposite chirality to Fig.~\ref{result1}. During the anticlockwise state-transfer $m_1\to m_2 \to m_3$, all the magnon modes achieve a nearly unit fidelity in a periodical sequence. Anticlockwise chiral state-transfer also applies to arbitrary states.

\begin{figure}[htbp]
\centering
\includegraphics[width=0.23\textwidth]{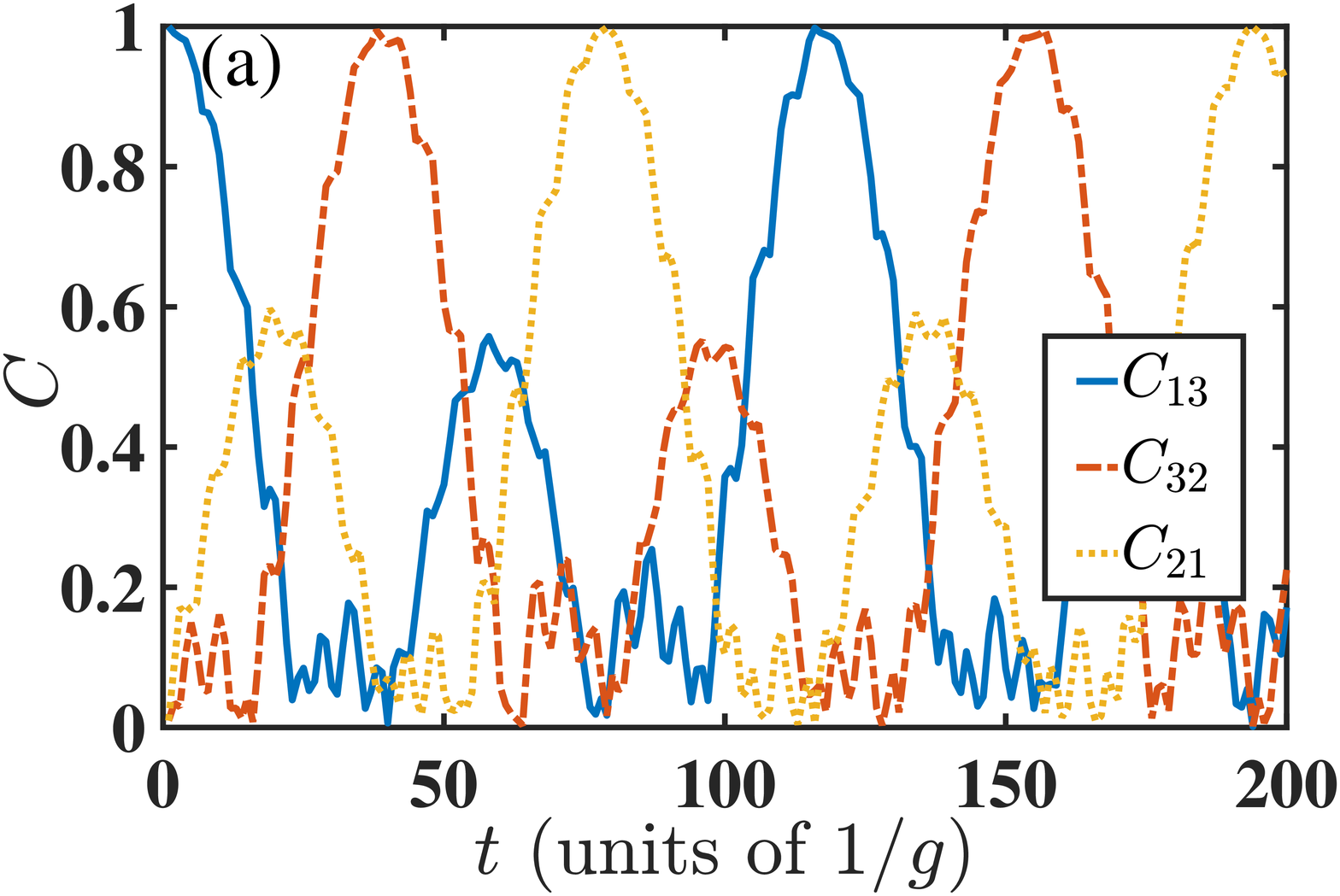}
\includegraphics[width=0.23\textwidth]{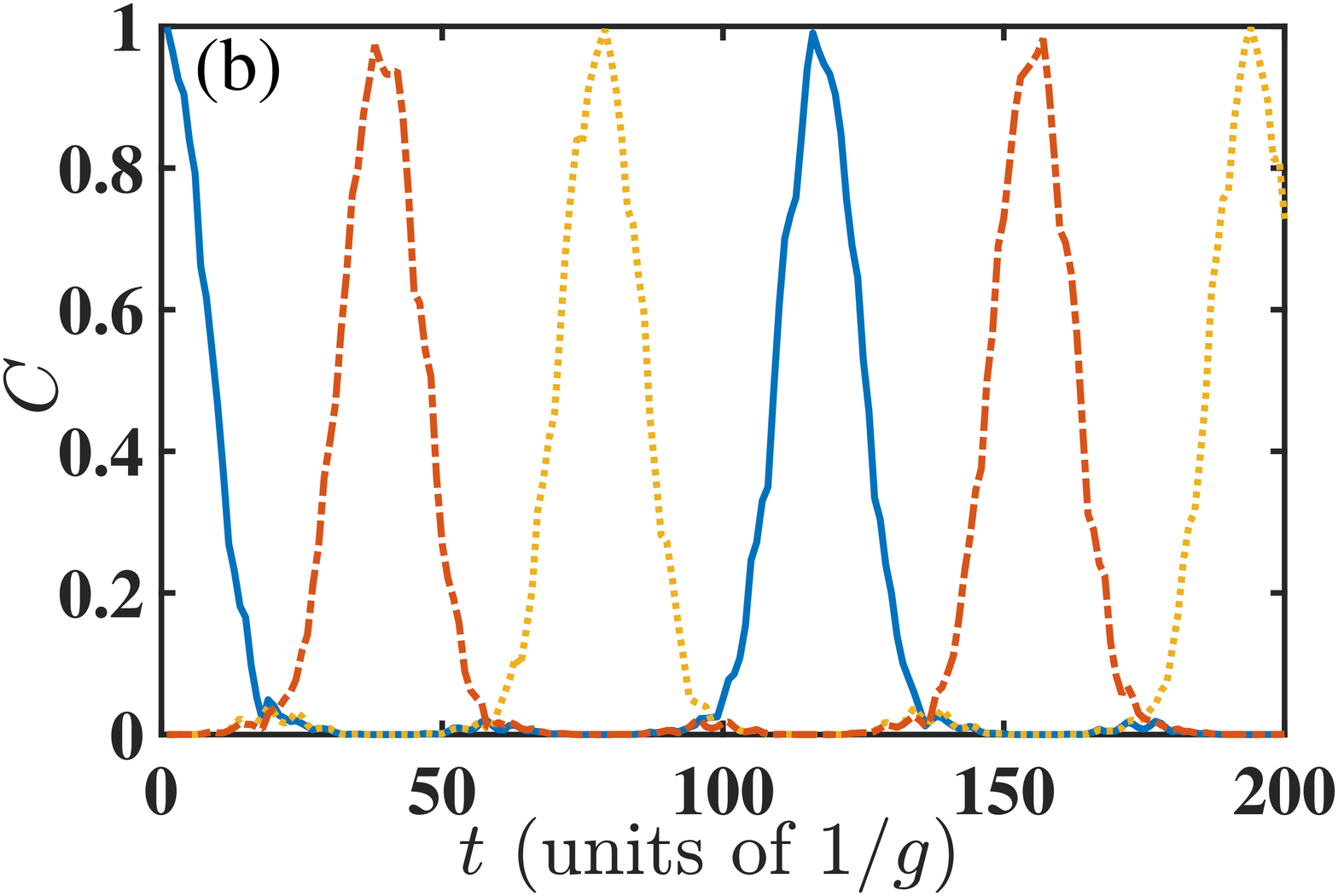}
\caption{Clockwise transfer of the pairwise concurrence in the three-magnon system under the Floquet-driving Hamiltonian~(\ref{Hamiltonian}). Magnonic mode-$1$ and mode-$3$ are initially set as (a) the Bell state $(|00\rangle+|11\rangle)/\sqrt{2}$ and (b) the NOON state $(|0N\rangle+|N0\rangle)/\sqrt{2}$ with $N=5$. The other parameters are the same as Fig.~\ref{anavsnum}.}\label{concur}
\end{figure}

If any two magnons in our model could be prepared as an entangled state and the third one is separably initialized as the ground state, we can demonstrate a high-fidelity chiral transfer for the pairwise quantum entanglement [see Fig.~\ref{model}(c)]. In particular, we consider that mode-$1$ and mode-$3$ are prepared as a double-excitation Bell state $(|00\rangle+|11\rangle)/\sqrt{2}$ or a NOON state $(|0N\rangle+|N0\rangle)/\sqrt{2}$ under the clockwise-transfer condition, i.e., $\Phi=\pi/2$.

It is well known that the concurrence could be used as a sufficient and necessary entanglement criterion for any bipartite system in pure state and any bipartite two-state system. According to its definition~\cite{concurrence} for a two-qubit system $AB$, $C=\max\{0, \lambda_1-\lambda_2-\lambda_3-\lambda_4\}$, where $\lambda_j$ are the square root of the eigenvalues of $\rho_{AB}(\sigma^A_y\otimes\sigma^B_y)\rho^*_{AB}(\sigma^A_y\otimes\sigma^B_y)$ in the decreasing order. For a pure state $a|11\rangle+b|10\rangle+c|01\rangle+d|00\rangle$ with normalized coefficients, its concurrence can be easily derived as $C=2|bc-ad|$.

Inspired by the expression for the pure state, we here employ a modified concurrence to measure the chiral-transfer of the entangled state in our model. If the initial state is the Bell state $(|00\rangle+|11\rangle)/\sqrt{2}$, then the concurrence between mode-$j$ and mode-$k$ is defined as $C_{jk}(t)=2|\langle\varphi(t)|00\rangle_{jk}||\langle\varphi(t)|11\rangle_{jk}|$; and if the target state is initially chosen as the NOON state $(|0N\rangle+|N0\rangle)/\sqrt{2}$, then $C_{jk}(t)=2|\langle\varphi(t)|0N\rangle_{jk}||\langle\varphi(t)|N0\rangle_{jk}|$. These definitions amount to the original occurrence under the perfect state transfer and become sufficient but not necessary for the entangled states during the time evolution, which is calculated with the Floquet-driving Hamiltonian~(\ref{Hamiltonian}).

In Fig.~\ref{concur}, the blue solid lines, the red dot-dashed lines, and the yellow dotted lines represent the concurrences for mode-$1$ and mode-$3$, mode-$3$ and mode-$2$, and mode-$2$ and mode-$1$, respectively. It is found that both the double-excitation Bell state and the NOON state could be transferred in a chiral way along the magnon triangle. At the same desired moments for the faithful state transfer, one can observe a nearly perfect entanglement transfer. Comparing Fig.~\ref{concur}(a) and Fig.~\ref{concur}(b), it is found that the undesired magnon pair obtains a lower probability of entanglement generation in transferring NOON state than Bell state. The side peaks of the Bell state $(|00\rangle+|11\rangle)/\sqrt{2}$ are dominant because its constituent $|00\rangle$ does not participate in the time evolution. Our model then widens the application range of the chiral state-transfer.

\subsection{Chiral current}\label{seccurrent}

Chiral current is an effective measure for the symmetry of the system Hamiltonian in condensed matter systems, which can be probed by measuring the energy spectrum under varying synthetic field $\Phi$~\cite{chiralcurrent}. Using the definition in Ref.~\cite{chiralcurrent}, we can consider the continuity equation for each magnon mode, e.g., $\partial_t\hat{n}_k=I_{\rm in}-I_{\rm out}$, where $\hat{n}_k\equiv m^{\dagger}_km_k$ is the occupation operator of the $k$th magnon mode and $I_{\rm in}$ ($I_{\rm out}$) represents the input (output) current with respect to mode-$k$. Under the necessary condition for the clockwise transfer in Fig.~\ref{model}(a), i.e., $\phi_j=2\pi j/3$, $j=1,2,3$, we have three equations of motion:
\begin{equation}\label{heisenop}
\begin{aligned}
\partial_t\hat{n}_1&=I_{12}-I_{31}=i\left[H_{\rm eff}, \hat{n}_1\right],\\
\partial_t\hat{n}_2&=I_{23}-I_{12}=i\left[H_{\rm eff}, \hat{n}_2\right],\\
\partial_t\hat{n}_3&=I_{31}-I_{23}=i\left[H_{\rm eff}, \hat{n}_3\right],\\
\end{aligned}
\end{equation}
where $H_{\rm eff}$ is the effective Hamiltonian in Eq.~(\ref{Heff0matrix}) or Eq.~(\ref{Heff0phi}) and $I_{jk}$ represents the current from mode-$j$ to mode-$k$. Then we have
\begin{equation}\label{curr}
\begin{aligned}
I_{12}=iGe^{if}m_1m^{\dagger}_2-iG^*e^{-if}m^{\dagger}_1m_2, \\
I_{31}=-iG^*e^{-if/2}m_1m^{\dagger}_3+iGe^{if/2}m^{\dagger}_1m_3, \\
I_{23}=iGe^{-if/2}m_2m^{\dagger}_3-iG^*e^{if/2}m^{\dagger}_2m_3. \\
\end{aligned}
\end{equation}
Normalized by the coupling-strength $|G|$ of $H_{\rm eff}$, the effective current operator from mode-$j$ to mode-$k$ can be defined as
\begin{equation}\label{currnor}
\hat{I}_{kj}=ie^{-i\phi_{kj}}m_km^{\dagger}_j-ie^{i\phi_{kj}}m^{\dagger}_km_j.
\end{equation}
Accordingly, the circle current operator is defined as $\hat{I}=\hat{I}_{12}+\hat{I}_{23}+\hat{I}_{31}$. It is straightforward to find that the circle current operator is an integral of motion because it commutates with the effective Hamiltonian, i.e., $[H_{\rm eff}, \hat{I}]=0$. It means that the expectation value of $\langle\hat{I}\rangle$ can not be used to measure the time-reversal symmetry of our model.

We have known that the magnon system has a symmetrical evolution for a synthetic magnetic flux $\Phi=0$ or $\Phi=\pi$ under the effective Hamiltonian in Eq.~(\ref{Heff0phi}). In this case, arbitrary state of mode-$1$ simultaneously propagates to mode-$2$ and mode-$3$ and periodically goes back to mode-$1$. Consequently, the current $\langle\hat{I}_{23}\rangle$ from mode-$2$ to mode-$3$ vanishes. A nonzero current $\langle\hat{I}_{23}\rangle$ appears when the time-reversal symmetry is broken when $\Phi\neq0, \pi$ and becomes maximum when $\Phi=\pm\pi/2$. We check the condition of $\Phi=\pi/2$ under $\phi_j=2\pi j/3$. Using $|\varphi(t)\rangle$ in Eq.~(\ref{timestate1}) with the initial Fock state $|100\rangle$, it is found that
\begin{equation}\label{current23}
\langle\hat{I}_{23}(t)\rangle=\frac{4}{9}\left[2\cos^2\left(\sqrt{3}g_{\rm eff}t\right)-\cos\left(\sqrt{3}g_{\rm eff}t\right)-1\right].
\end{equation}
The maximal value is achieved as $\langle\hat{I}_{23}^{\rm max}\rangle=8/9$ when $\sqrt{3}g_{\rm eff}t=\pi+2n\pi$. Given an arbitrary state in Eq.~(\ref{statet}), we have
\begin{equation}\label{currentmax}
\langle\hat{I}_{23}^{\rm max}\rangle=\frac{8}{9}\sum_nnC^2_n=\frac{8}{9}\bar{n},
\end{equation}
where $\bar{n}$ is the average excitation number of the state. For a coherent state $|\beta\rangle$, $\bar{n}=|\beta|^2$ and for a cat state $(|\zeta\rangle+|-\zeta\rangle)/\sqrt{2+2e^{-2|\zeta|^2}}$, $\bar{n}=|\zeta|^2[1-\exp(-2|\zeta|^2)]/[1+\exp(-2|\zeta|^2)]$. Then the expectation value of $\langle\hat{I}_{23}(t)\rangle$ is effective to measure the time-reversal or parity-reversal symmetry of the system Hamiltonian.

\begin{figure}[htbp]
\centering
\includegraphics[width=0.23\textwidth]{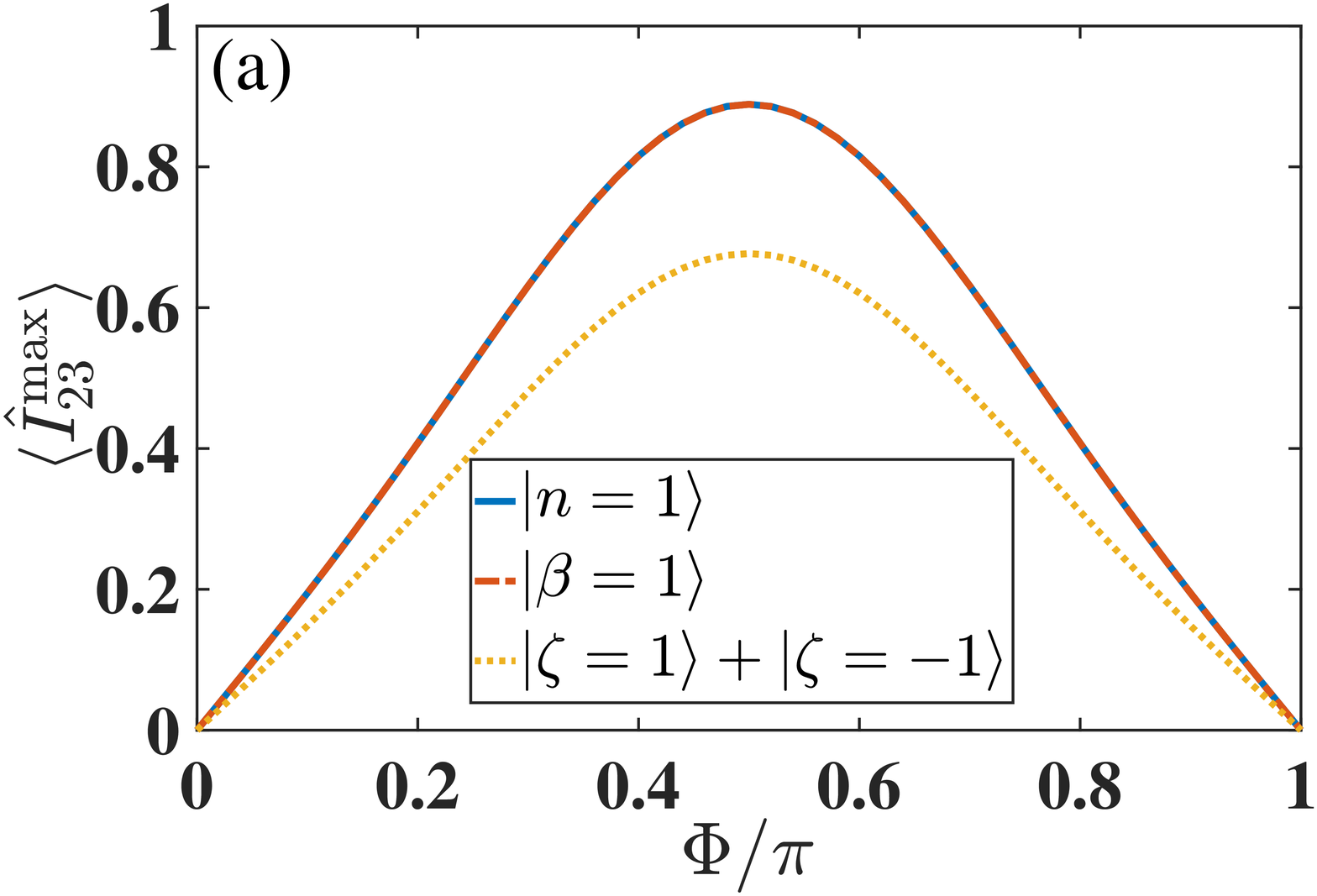}
\includegraphics[width=0.23\textwidth]{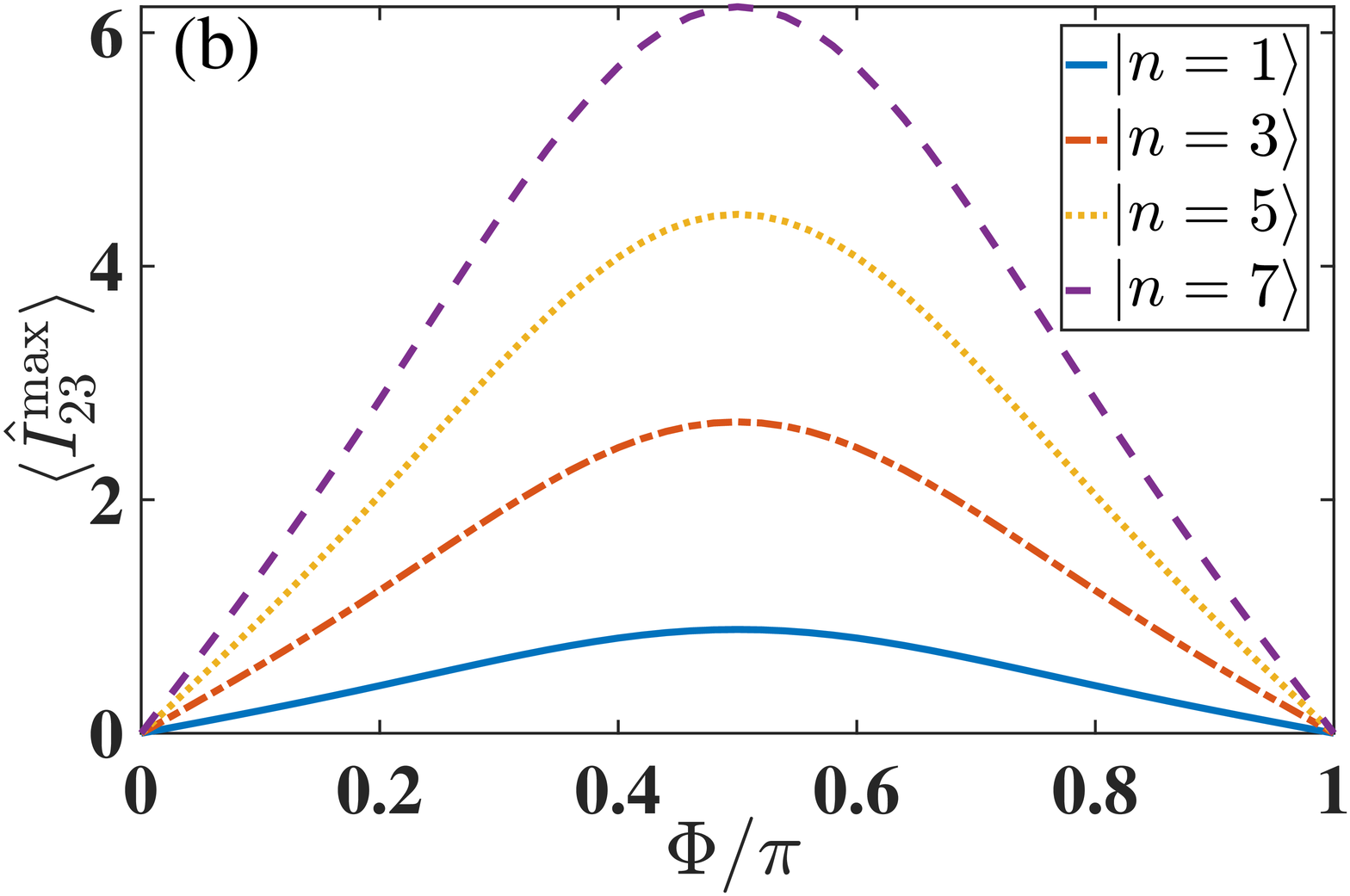}
\caption{(a) Maximal value of the current from $m_2$ to $m_3$ as a function of the closed-loop phase $\Phi$ under the effective Hamiltonian~(\ref{Heff0phi}). The initial states for $m_1$ are the Fock state $|n=1\rangle$ (the blue solid line),  the coherent state with $\beta=1$ (the red dot-dashed line), and the cat state with $\zeta=1$ (the yellow dotted line). (b) Maximal value of the current from $m_2$ to $m_3$ for various Fock states as a function of $\Phi$. }\label{current1}
\end{figure}

The preceding analysis is consistent with the numerical simulation in Fig.~\ref{current1} about $\langle\hat{I}_{23}^{\rm max}\rangle$ as a function of the close-loop phase or the synthetic magnetic field $\Phi$. It is found that the maximal current vanishes when $\Phi=0$ and $\Phi=\pi$. It can be understood that at these points, the system Hamiltonian becomes real that cannot break the time-reversal symmetry. Otherwise, any finite $\Phi$ breaks the symmetry and leads to the chiral currents. It is interesting to find in Fig.~\ref{current1}(a) that the maximal currents for the Fock state $|n=1\rangle$ and the coherent state with $\beta=1$ are equivalent to each other. Both of them equal to $0.89\approx8/9$ at the peak point $\Phi=\pi/2$. They are larger than the cat state with $\zeta=1$, whose peak value is $0.68$. A larger average excitation number yields a larger $\langle\hat{I}_{23}^{\rm max}\rangle$ with the same $\Phi$. This result is also supported by Fig.~\ref{current1}(b), in which a monotonic pattern appears with increasing $n$.

\section{Fidelity analysis}

\subsection{Fidelity analysis under magnon damping}\label{damping}

Taking account the magnon damping into consideration, we can study the chiral state-transfer fidelity in a standard open-quantum-system framework. In general, the full Hamiltonian including the environments can be written as
\begin{equation}\label{open}
H_{\rm tot}=H(t)+\sum_{k=1}^3\left(m_kB_k^\dagger+m_k^\dagger B_k\right),
\end{equation}
where $H(t)$ is the Floquet-driving Hamiltonian~(\ref{Hamiltonian}) with $N=3$ and $B_k=\sum_jg_{kj}a^\dagger_{kj}e^{-i\omega_{kj}t}$ is a collective environmental operator. $\omega_{kj}$ is the eigenfrequency for the $j$th field mode in the $k$th environment and $g_{kj}$ is its coupling strength with the $k$th magnon mode. Under the Markovian approximation and tracing out the degrees of freedom of the external environments (assumed to be at the vacuum state), we arrive at the master equation~\cite{opentheory} for the density-matrix operator $\rho(t)$ for the magnon system:
\begin{equation}\label{mastereq}
\dot{\rho}(t)=-i\left[H(t), \rho(t)\right]+\kappa\sum_{k=1}^3\mathcal{L}[m_k]\rho(t), \\
\end{equation}
where $\kappa$ is the damping rate. The dissipative superoperator $\mathcal{L}$ is in a Lindblad form,
\begin{equation}\label{operator}
\mathcal{L}[o]\rho\equiv\frac{1}{2}\left(2o\rho o^\dag-o^\dag o\rho-\rho o^\dag o\right),
\end{equation}
where $o=m_1,m_2,m_3$ are the decay operators.

\begin{figure}[htbp]
\centering
\includegraphics[width=0.23\textwidth]{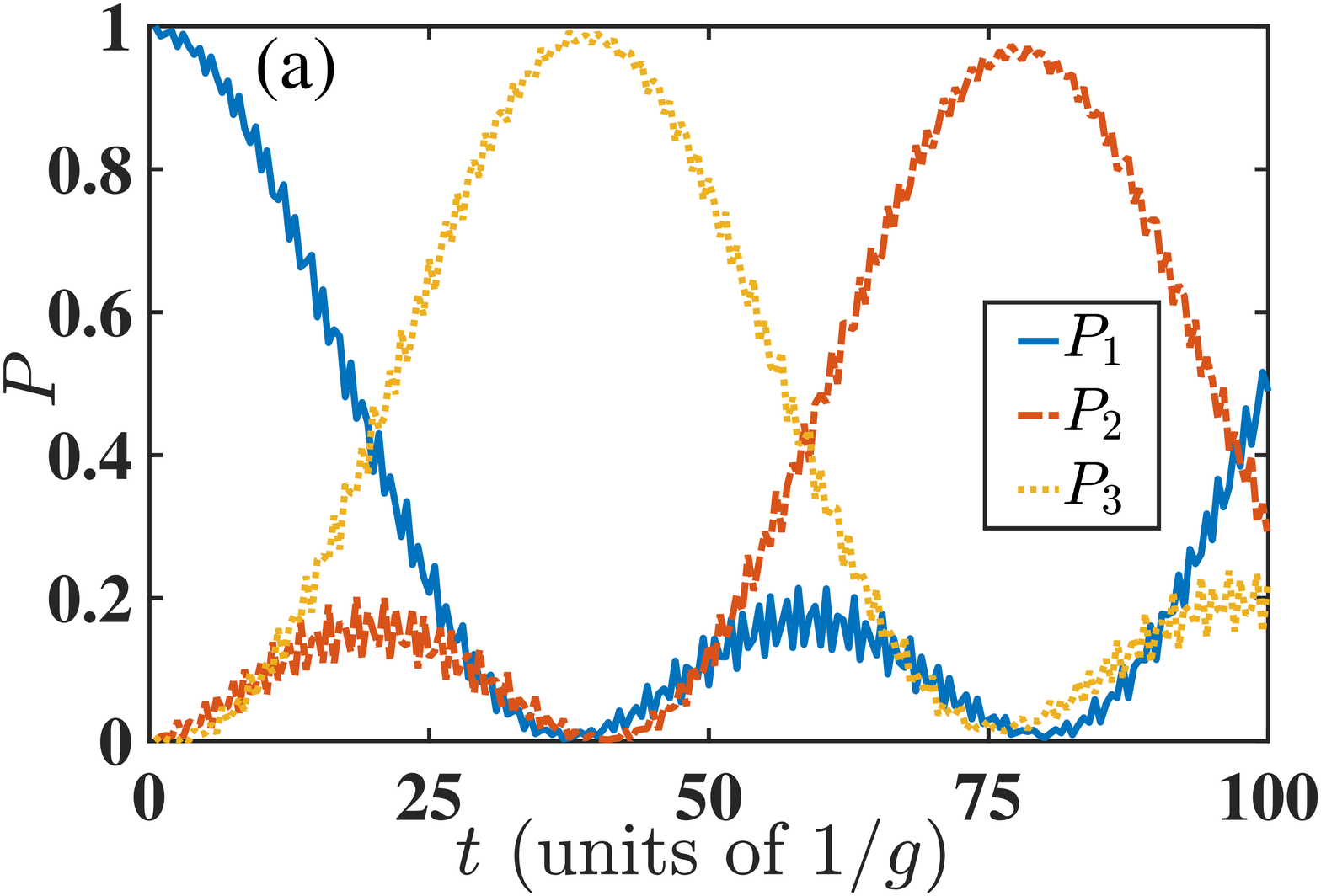}
\includegraphics[width=0.23\textwidth]{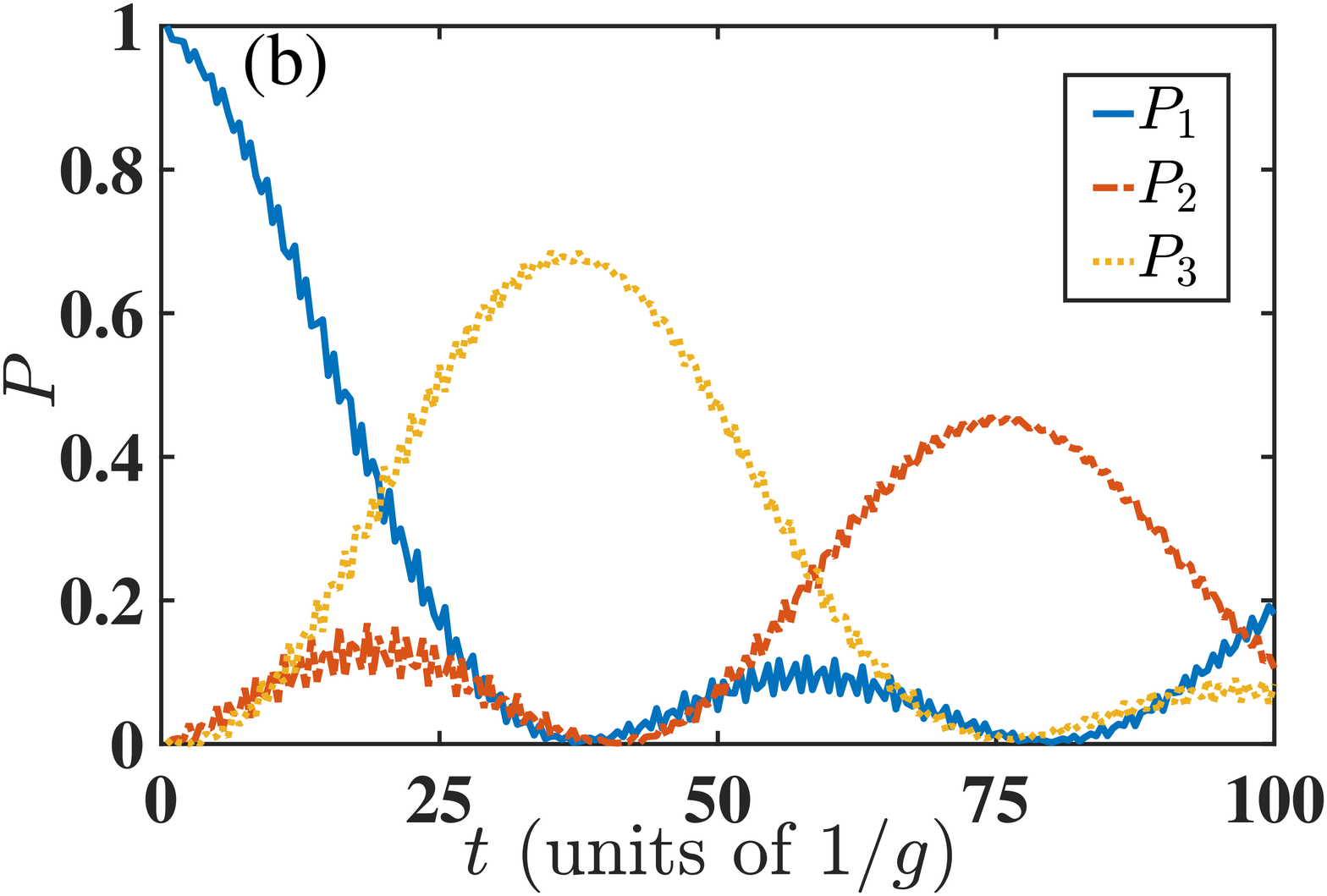}
\caption{Clockwise state-transfer of the three magnons described by $P_j$'s under various damping rates: (a) $\kappa/g=0.001$ and (b) $\kappa/g=0.01$. The initial state is $|n=1\rangle|0\rangle|0\rangle$. $\omega=20g$ and $\phi_j=2\pi j/3$. }\label{master}
\end{figure}

In Fig.~\ref{master}(a), one can observe that the open-system dynamics under $\kappa/g=0.001$ is very close to the closed-system result in Fig.~\ref{result1}(a). And in Fig.~\ref{master}(b), it is found that at least one round of chiral transfer presents when $\kappa/g=0.01$. It is thus required that the effective coupling strength $g$ is about two orders higher than the damping rate in magnitude. For YIG spheres of a typical size about $250$ $\mu$m in recent experiments~\cite{yigcavity,magnonqubit,magnonqubit2}, the coupling strength $g$ was found to be in the order of $1\sim 10$ MHz and the damping rate $\kappa\sim 1$ MHz. Then our protocol could become practical in a near future by increasing the strength of the magnetic dipole interaction to the order of $100$ MHz~\cite{magnon}.

Alternatively, we can apply the external quantum control, such as the dynamical decoupling method~\cite{suppre}, to our system to neutralize the magnon damping effects. The system coherence can then be saved by the $\pi$-pulses exerted on the magnons. It is known that
\begin{equation}
e^{i\int_{\tau_n}^{\tau_{n+1}}dt\Omega(t)m_k^\dagger m_k}m_ke^{-i\int_{\tau_n}^{\tau_{n+1}}dt\Omega(t)m_k^\dagger m_k}=-m_k
\end{equation}
as long as $\int_{\tau_n}^{\tau_{n+1}}dt\Omega(t)=\pi$. In the ideal situation for dynamical decoupling, $\Omega(t)=\pi/(\tau_{n+1}-\tau_n)\delta(t-s)$ with $s\in(\tau_n, \tau_{n+1})$. Thus it could reverse the decoherence dynamics induced by the unwanted terms $m_kB_k^\dagger+H.c.$ in Eq.~(\ref{open}), but does not change the effective Hamiltonian in our protocol.

\subsection{Fidelity under systematic errors}\label{syserror}

Our protocol based on Eq.~(\ref{Ham}) has been discussed under the ideal condition that all the coupling strengths $g_{am}$ between photon and magnon modes are the same in magnitude. While in practice~\cite{yigcavity2,magnon}, they are associated with the individual locations of the YIG spheres in the cavity. We first consider the systematic errors raised by the nonequal coupling strengths. In particular, it is assumed that $g_{am}^{(1)}=g_{am}$, $g_{am}^{(2)}=g_{am}(1+\delta)$, and $g_{am}^{(3)}=g_{am}(1-\delta)$, where the superscript $j\in\{1,2,3\}$ marks the coupling strength between the common photon and the $j$th magnon and $\delta$ is used to estimate the magnitude of the relative error. With $N=3$, the system Hamiltonian in Eq.~(\ref{Ham}) is then rewritten as
\begin{equation}\label{Hamdelta}
\begin{aligned}
H&=\omega_a a^{\dagger}a+\omega_m\sum_{k=1}^3m^{\dagger}_km_k+g_{am}\left(am^{\dagger}_1+a^{\dagger}m_1\right)\\
&+ g_{am}(1+\delta)\left(am^{\dagger}_2+a^{\dagger}m_2\right)\\&+ g_{am}(1-\delta)\left(am^{\dagger}_3+a^{\dagger}m_3\right).
\end{aligned}
\end{equation}

Under the conditions that $\phi_j=2\pi j/3$ and $J_0(f)=0$ for the clockwise chiral transfer, the effective Hamiltonian in Eq.~(\ref{Heffmatrix}) is modified by changing the coefficient matrix to be
\begin{equation}\label{Heffmatrixdelta}
ig_{\rm eff}\begin{bmatrix}
0 & -(1-\delta^2)e^{if} & (1+\delta)e^{if/2}\\
(1-\delta^2)e^{-if} & 0 & -(1-\delta)e^{-if/2}\\
-(1+\delta)e^{-if/2} & (1-\delta)e^{if/2} & 0
\end{bmatrix}.
\end{equation}

\begin{figure}[htbp]
\centering
\includegraphics[width=0.23\textwidth]{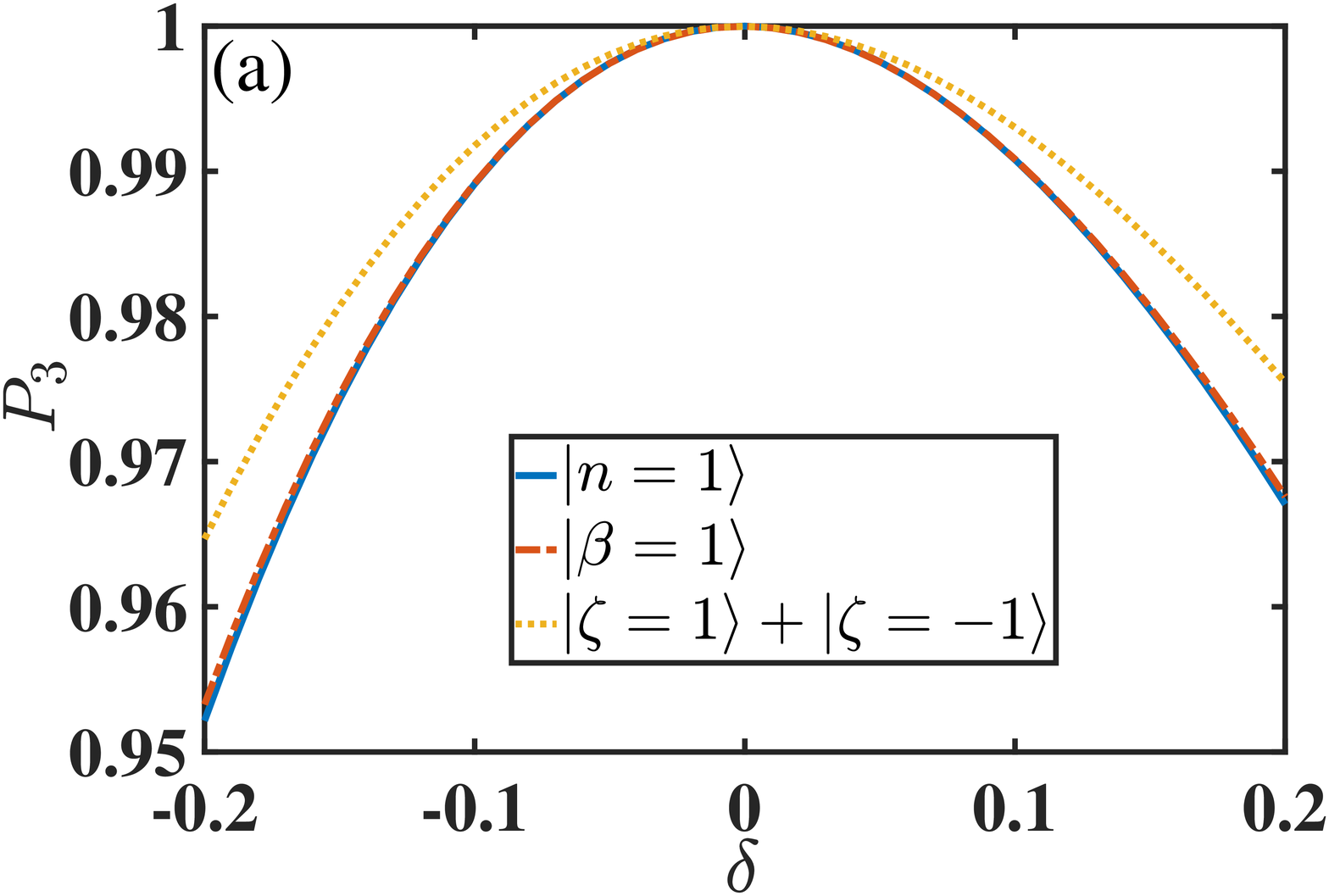}
\includegraphics[width=0.23\textwidth]{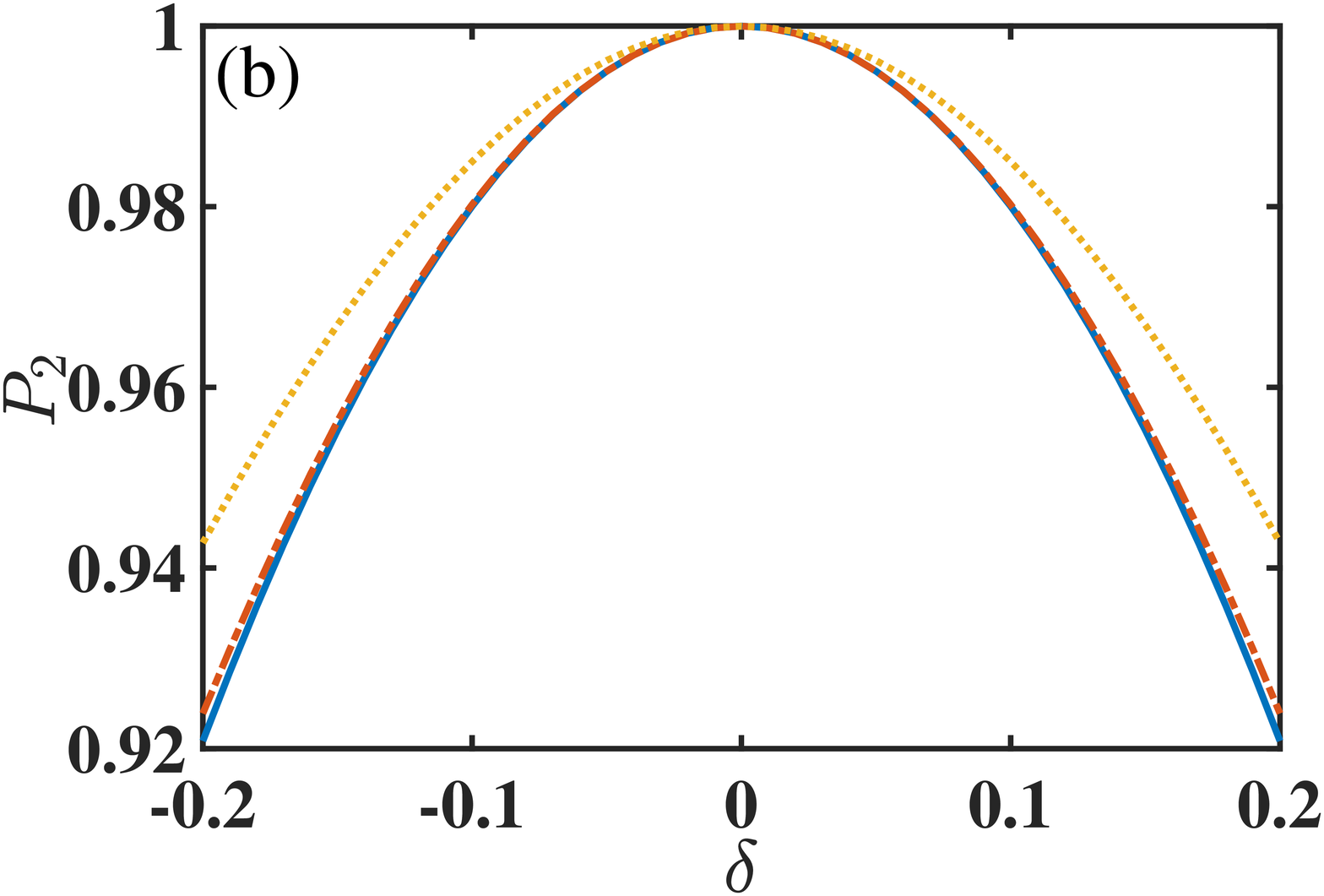}
\caption{(a) State population of the $3$rd magnon at the desired moment $t_3=2\pi/(3\sqrt{3}g_{\rm eff})$ and (b) State population of the $2$nd magnon at the desired moment $t_2=4\pi/(3\sqrt{3}g_{\rm eff})$ as functions of the magnitude of the relative error $\delta$ under the nonideal Hamiltonian~(\ref{Heffmatrixdelta}). The initial states for $m_1$ are the Fock state $|n=1\rangle$ (the blue solid line),  the coherent state with $\beta=1$ (the red dot-dashed line), and the cat state with $\zeta=1$ (the yellow dotted line). }\label{error}
\end{figure}

In Fig.~\ref{error}, we present the sensitivity of the state populations [see Eq.~(\ref{population})] to the error $\delta$ for $g_{ma}$ at the desired moments for the chiral state-transfer obtained by Eq.~(\ref{xyz}). It is found that the robustness of our protocol against the systematic errors becomes weaker with a larger average population. The cat state with $\zeta=1$ is found to be more robust in comparison with the Fock state $|n=1\rangle$ and the coherent state with $\beta=1$. For all of them, both transferred populations of mode-$2$ and mode-$3$ could be maintained above $0.98$ in the presence of about $10\%$ errors or fluctuations in the photon-magnon coupling strength.

Another systematic error in our model is associated with the nonlinearity in the expansion of the YIG crystal or the dipolar anisotropy energy quadratic in the magnon numbers, which is parameterized with the Kerr coefficient $K$. When the excitation number in the spin wave is much smaller than the total number of spins in YIG sphere, the system Hamiltonian in Eq.~(\ref{Heffsecond}) can be modified to~\cite{kerrmagnon,kerrmagnon2}
\begin{equation}\label{Heffsecondkerr}
\tilde{H}_{\rm Kerr}=g\sum_{k<j}\left(m_km^{\dagger}_j+m^{\dagger}_km_j\right)+K\sum_{k}\left(m^{\dagger}_km_k\right)^2.
\end{equation}
The Kerr coefficient $K/2\pi\approx 6.4\times 10^{-9}$ Hz, which is much smaller than the effective coupling strength $g$. Given the timescale we considered in the preceding chiral dynamics, we can safely apply our model in such a hybrid photon-magnon system when the number of the excited spins $\langle m^{\dagger}_km_k\rangle$ is less than $10^6$.

\section{Conclusion}\label{conclu}

In summary, we have proposed a chiral state-transfer proposal in a Floquet cavity-magnonic system, which consists of a microwave cavity coupled to three YIG spheres in their Kittle mode. The three magnon modes constitute a triangle loop of mutual interaction through eliminating the photon mode in a state-resolved way. The Floquet driving with desired parameters forms complex-valued hopping of magnons, that generates a tunable synthetic magnetic flux. Our results open a path toward simulation of a time-reversal-broken continuous-variable system, yielding the chiral transfer for arbitrary states and even quantum entanglement. It is beyond the limitation of fixed excitation numbers. We evaluate the chiral current to measure the symmetry of the system Hamiltonian. We discuss the feasibility of the chiral transfer in the presence of magnon damping. The stability of our proposal is also tested with respect to the systematic errors in the coupling strength and the nonlinear terms in the system. Our work in pursuit of the quantum chiral state transfer provides an interesting application in cavity magnonics as a promising hybrid platform for control over the emerging phases.

Our theory can be straightforwardly generalized to arbitrary continuous-variable systems that could be indirectly coupled to each other by a common mediator. For example, one can replace our magnon modes with microwave photon modes in superconducting circuit with a much reduced ratio of decoherence and coupling strength~\cite{cq}. In addition, although our protocol has no direct relation with topological variables, the anomalous edge states, and the bulk-edge correspondence in the momentum space, we expect that the edge state~\cite{topolight,chiralcurrent,edgestate} can arise from the cavity-magnonic system of a larger size under Floquet engineering. It will bring extra interesting physics.

\section*{Acknowledgments}

We acknowledge financial support from the National Science Foundation of China (Grants No. 11974311 and No. U1801661).

\appendix

\section{RWA in Hamiltonian~(\ref{Ham})}\label{appa}

This appendix contributes to verifying that it is valid to apply the rotating-wave approximation in the model Hamiltonian~(\ref{Ham}). Without loss of generality, it is instructive to focus on the state exchange (Rabi oscillation) between two magnons mediated by a common photon mode as what we concerned in Eq.~(\ref{Hsub}). Including the counter-rotating terms, the Hamiltonian reads
\begin{equation}\label{Htot}
H=\omega_aa^{\dagger}a+\omega_m\sum_{k=1}^2m^{\dagger}_km_k+g_{am}\left(a+a^{\dagger}\right)\sum_{k=1}^2 \left(m^{\dagger}_k+m_k\right).
\end{equation}
And it becomes
\begin{equation}\label{Hrwa}
H_{\rm rwa}=\omega_aa^{\dagger}a+\omega_m\sum_{k=1}^2m^{\dagger}_km_k+g_{am}\sum_{k=1}^2 \left(am^{\dagger}_k+a^{\dagger}m_k\right)
\end{equation}
under RWA.

\begin{figure}[htbp]
\centering
\includegraphics[width=0.23\textwidth]{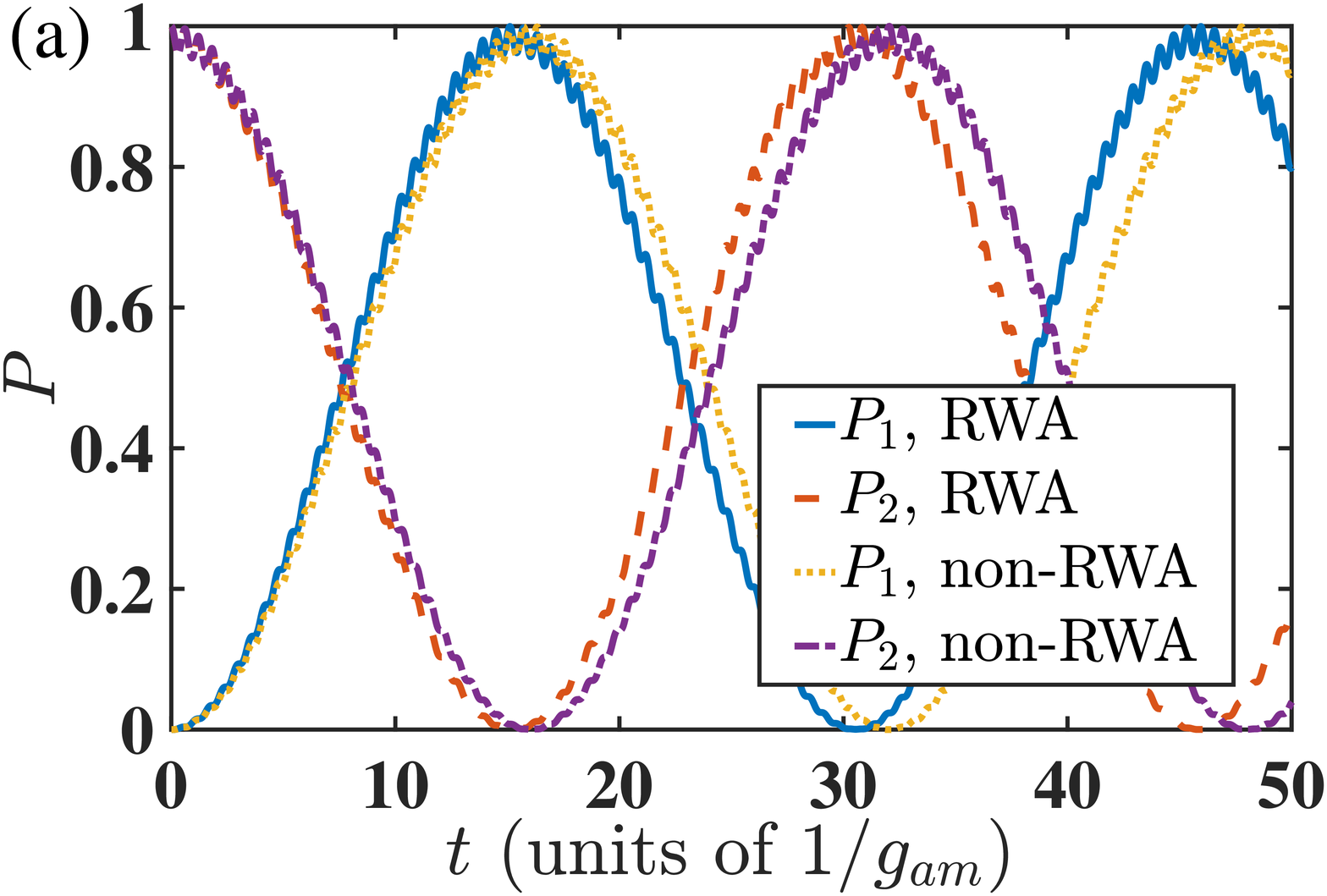}
\includegraphics[width=0.23\textwidth]{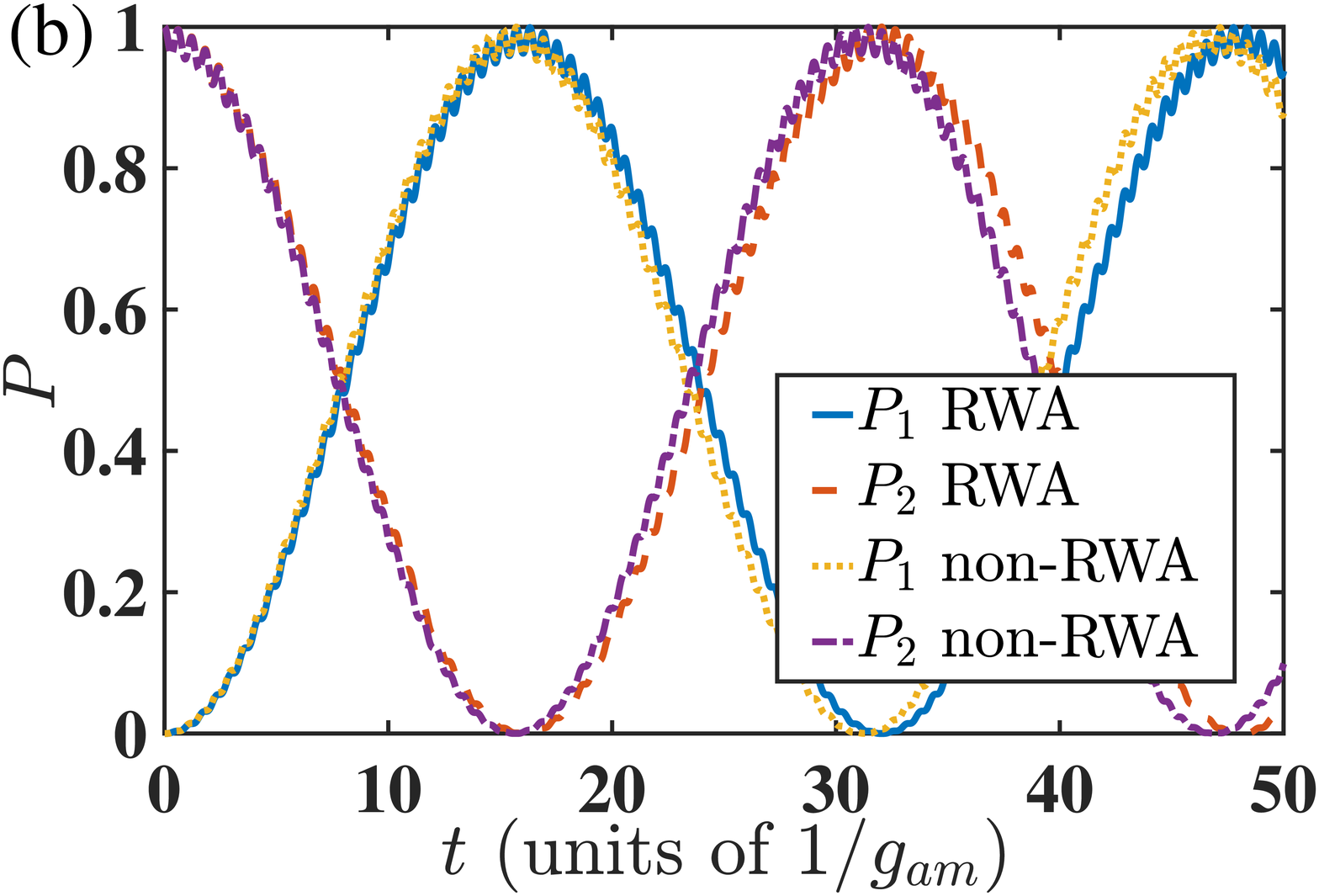}
\caption{Time evolution of the state populations $P(t)$ for the two magnon modes evaluated by the Hamiltonian in Eq.~(\ref{Htot}) and that under RWA in Eq.~(\ref{Hrwa}). The initial state is $|\varphi(0)\rangle=|1\rangle_{m_1}|0\rangle_{m_2}$ and the detuning is chosen as $\Delta=10g_{am}$. In (a), $\omega_m=100g_{am}$ and in (b), $\omega_m=200g_{am}$. }\label{counter}
\end{figure}

To demonstrate the effect from the counter-rotating terms in a practical cavity-magnon system~\cite{kerrmagnon,kerrmagnon2} on Rabi oscillation of the two modes, we plot their population dynamics using the total Hamiltonian in Eq.~(\ref{Htot}) and that under RWA in Eq.~(\ref{Hrwa}). One can find that in Fig.~\ref{counter}(a), the difference between $H$ and $H_{\rm rwa}$ is not significant when $\omega_m=100g_{am}$; and in Fig.~\ref{counter}(b) it becomes ignorable when $\omega_m=200g_{am}$. Thus we are sure that it is not necessary to include the counter-rotating terms since $g_{am}\ll\Delta=\omega_a-\omega_m$ is in the strong instead of ultrastrong regimes.

Note the counter-rotating terms does not constitute an obstacle for our state-resolved method~\cite{intermagnon} to eliminate the mediator [see from Eq.~(\ref{Hsub}) to Eq.~(\ref{tildeH})]. We can still obtain an effective Hamiltonian for the magnons as in Eq.~(\ref{tildeH}), and it is little different from the current result in the dispersive regime of $g_{am}$. 

\bibliographystyle{apsrevlong}
\bibliography{reference}

\begin{thebibliography}{61}%
\makeatletter
\providecommand \@ifxundefined [1]{%
 \@ifx{#1\undefined}
}%
\providecommand \@ifnum [1]{%
 \ifnum #1\expandafter \@firstoftwo
 \else \expandafter \@secondoftwo
 \fi
}%
\providecommand \@ifx [1]{%
 \ifx #1\expandafter \@firstoftwo
 \else \expandafter \@secondoftwo
 \fi
}%
\providecommand \natexlab [1]{#1}%
\providecommand \enquote  [1]{``#1''}%
\providecommand \bibnamefont  [1]{#1}%
\providecommand \bibfnamefont [1]{#1}%
\providecommand \citenamefont [1]{#1}%
\providecommand \href@noop [0]{\@secondoftwo}%
\providecommand \href [0]{\begingroup \@sanitize@url \@href}%
\providecommand \@href[1]{\@@startlink{#1}\@@href}%
\providecommand \@@href[1]{\endgroup#1\@@endlink}%
\providecommand \@sanitize@url [0]{\catcode `\\12\catcode `\$12\catcode
  `\&12\catcode `\#12\catcode `\^12\catcode `\_12\catcode `\%12\relax}%
\providecommand \@@startlink[1]{}%
\providecommand \@@endlink[0]{}%
\providecommand \url  [0]{\begingroup\@sanitize@url \@url }%
\providecommand \@url [1]{\endgroup\@href {#1}{\urlprefix }}%
\providecommand \urlprefix  [0]{URL }%
\providecommand \Eprint [0]{\href }%
\providecommand \doibase [0]{http://dx.doi.org/}%
\providecommand \selectlanguage [0]{\@gobble}%
\providecommand \bibinfo  [0]{\@secondoftwo}%
\providecommand \bibfield  [0]{\@secondoftwo}%
\providecommand \translation [1]{[#1]}%
\providecommand \BibitemOpen [0]{}%
\providecommand \bibitemStop [0]{}%
\providecommand \bibitemNoStop [0]{.\EOS\space}%
\providecommand \EOS [0]{\spacefactor3000\relax}%
\providecommand \BibitemShut  [1]{\csname bibitem#1\endcsname}%
\let\auto@bib@innerbib\@empty
\bibitem [{\citenamefont {Rameshti}\ \emph {et~al.}(2021)\citenamefont
  {Rameshti}, \citenamefont {Kusminskiy}, \citenamefont {Haigh}, \citenamefont
  {Usami}, \citenamefont {Lachance-Quirion}, \citenamefont {Nakamura},
  \citenamefont {Hu}, \citenamefont {Tang}, \citenamefont {Bauer},\ and\
  \citenamefont {Blanter}}]{cavitymagnonics}%
  \BibitemOpen
  \bibfield  {author} {\bibinfo {author} {\bibfnamefont {B.~Z.}\ \bibnamefont
  {Rameshti}}, \bibinfo {author} {\bibfnamefont {S.~V.}\ \bibnamefont
  {Kusminskiy}}, \bibinfo {author} {\bibfnamefont {J.~A.}\ \bibnamefont
  {Haigh}}, \bibinfo {author} {\bibfnamefont {K.}~\bibnamefont {Usami}},
  \bibinfo {author} {\bibfnamefont {D.}~\bibnamefont {Lachance-Quirion}},
  \bibinfo {author} {\bibfnamefont {Y.}~\bibnamefont {Nakamura}}, \bibinfo
  {author} {\bibfnamefont {C.-M.}\ \bibnamefont {Hu}}, \bibinfo {author}
  {\bibfnamefont {H.~X.}\ \bibnamefont {Tang}}, \bibinfo {author}
  {\bibfnamefont {G.~E.}\ \bibnamefont {Bauer}}, \ and\ \bibinfo {author}
  {\bibfnamefont {Y.~M.}\ \bibnamefont {Blanter}},\ }\bibfield  {title} {\emph
  {\bibinfo {title} {Cavity magnonics},\ }}\href@noop {} {\bibfield  {journal}
  {\bibinfo  {journal} {arxiv: 2106. 09312v1}\ } (\bibinfo {year}
  {2021})}\BibitemShut {NoStop}%
\bibitem [{\citenamefont {Lachance-Quirion}\ \emph {et~al.}(2019)\citenamefont
  {Lachance-Quirion}, \citenamefont {Tabuchi}, \citenamefont {Gloppe},
  \citenamefont {Usami},\ and\ \citenamefont {Nakamura}}]{magnon}%
  \BibitemOpen
  \bibfield  {author} {\bibinfo {author} {\bibfnamefont {D.}~\bibnamefont
  {Lachance-Quirion}}, \bibinfo {author} {\bibfnamefont {Y.}~\bibnamefont
  {Tabuchi}}, \bibinfo {author} {\bibfnamefont {A.}~\bibnamefont {Gloppe}},
  \bibinfo {author} {\bibfnamefont {K.}~\bibnamefont {Usami}}, \ and\ \bibinfo
  {author} {\bibfnamefont {Y.}~\bibnamefont {Nakamura}},\ }\bibfield  {title}
  {\emph {\bibinfo {title} {Hybrid quantum systems based on magnonics},\
  }}\href {\doibase 10.7567/1882-0786/ab248d} {\bibfield  {journal} {\bibinfo
  {journal} {Appl. Phys. Express}\ }\textbf {\bibinfo {volume} {12}},\ \bibinfo
  {pages} {070101} (\bibinfo {year} {2019})}\BibitemShut {NoStop}%
\bibitem [{\citenamefont {Li}\ \emph {et~al.}(2020)\citenamefont {Li},
  \citenamefont {Zhang}, \citenamefont {Tyberkevych}, \citenamefont {Kwok},\
  and\ \citenamefont {Novosad}}]{magnon2}%
  \BibitemOpen
  \bibfield  {author} {\bibinfo {author} {\bibfnamefont {Y.}~\bibnamefont
  {Li}}, \bibinfo {author} {\bibfnamefont {W.}~\bibnamefont {Zhang}}, \bibinfo
  {author} {\bibfnamefont {V.}~\bibnamefont {Tyberkevych}}, \bibinfo {author}
  {\bibfnamefont {W.~K.}\ \bibnamefont {Kwok}}, \ and\ \bibinfo {author}
  {\bibfnamefont {V.}~\bibnamefont {Novosad}},\ }\bibfield  {title} {\emph
  {\bibinfo {title} {Hybrid magnonics: physics, circuits, and applications for
  coherent information processing},\ }}\href {\doibase 10.1063/5.0020277}
  {\bibfield  {journal} {\bibinfo  {journal} {J. Appl. Phys.}\ }\textbf
  {\bibinfo {volume} {128}},\ \bibinfo {pages} {130902} (\bibinfo {year}
  {2020})}\BibitemShut {NoStop}%
\bibitem [{\citenamefont {Yuan}\ \emph {et~al.}(2022)\citenamefont {Yuan},
  \citenamefont {Cao}, \citenamefont {Kamra}, \citenamefont {Duine},\ and\
  \citenamefont {Yan}}]{magnonstate}%
  \BibitemOpen
  \bibfield  {author} {\bibinfo {author} {\bibfnamefont {H.}~\bibnamefont
  {Yuan}}, \bibinfo {author} {\bibfnamefont {Y.}~\bibnamefont {Cao}}, \bibinfo
  {author} {\bibfnamefont {A.}~\bibnamefont {Kamra}}, \bibinfo {author}
  {\bibfnamefont {R.~A.}\ \bibnamefont {Duine}}, \ and\ \bibinfo {author}
  {\bibfnamefont {P.}~\bibnamefont {Yan}},\ }\bibfield  {title} {\emph
  {\bibinfo {title} {Quantum magnonics: When magnon spintronics meets quantum
  information science},\ }}\href {\doibase
  https://doi.org/10.1016/j.physrep.2022.03.002} {\bibfield  {journal}
  {\bibinfo  {journal} {Phys. Rep.}\ }\textbf {\bibinfo {volume} {965}},\
  \bibinfo {pages} {1} (\bibinfo {year} {2022})}\BibitemShut {NoStop}%
\bibitem [{\citenamefont {Soykal}\ and\ \citenamefont
  {Flatt\'e}(2010{\natexlab{a}})}]{cavityyig1}%
  \BibitemOpen
  \bibfield  {author} {\bibinfo {author} {\bibfnamefont {O.~O.}\ \bibnamefont
  {Soykal}}\ and\ \bibinfo {author} {\bibfnamefont {M.~E.}\ \bibnamefont
  {Flatt\'e}},\ }\bibfield  {title} {\emph {\bibinfo {title} {Strong field
  interactions between a nanomagnet and a photonic cavity},\ }}\href {\doibase
  10.1103/PhysRevLett.104.077202} {\bibfield  {journal} {\bibinfo  {journal}
  {Phys. Rev. Lett.}\ }\textbf {\bibinfo {volume} {104}},\ \bibinfo {pages}
  {077202} (\bibinfo {year} {2010}{\natexlab{a}})}\BibitemShut {NoStop}%
\bibitem [{\citenamefont {Soykal}\ and\ \citenamefont
  {Flatt\'e}(2010{\natexlab{b}})}]{cavityyig2}%
  \BibitemOpen
  \bibfield  {author} {\bibinfo {author} {\bibfnamefont {O.~O.}\ \bibnamefont
  {Soykal}}\ and\ \bibinfo {author} {\bibfnamefont {M.~E.}\ \bibnamefont
  {Flatt\'e}},\ }\bibfield  {title} {\emph {\bibinfo {title} {Size dependence
  of strong coupling between nanomagnets and photonic cavities},\ }}\href
  {\doibase 10.1103/PhysRevB.82.104413} {\bibfield  {journal} {\bibinfo
  {journal} {Phys. Rev. B}\ }\textbf {\bibinfo {volume} {82}},\ \bibinfo
  {pages} {104413} (\bibinfo {year} {2010}{\natexlab{b}})}\BibitemShut
  {NoStop}%
\bibitem [{\citenamefont {Tabuchi}\ \emph {et~al.}(2014)\citenamefont
  {Tabuchi}, \citenamefont {Ishino}, \citenamefont {Ishikawa}, \citenamefont
  {Yamazaki}, \citenamefont {Usami},\ and\ \citenamefont
  {Nakamura}}]{yigcavity}%
  \BibitemOpen
  \bibfield  {author} {\bibinfo {author} {\bibfnamefont {Y.}~\bibnamefont
  {Tabuchi}}, \bibinfo {author} {\bibfnamefont {S.}~\bibnamefont {Ishino}},
  \bibinfo {author} {\bibfnamefont {T.}~\bibnamefont {Ishikawa}}, \bibinfo
  {author} {\bibfnamefont {R.}~\bibnamefont {Yamazaki}}, \bibinfo {author}
  {\bibfnamefont {K.}~\bibnamefont {Usami}}, \ and\ \bibinfo {author}
  {\bibfnamefont {Y.}~\bibnamefont {Nakamura}},\ }\bibfield  {title} {\emph
  {\bibinfo {title} {Hybridizing ferromagnetic magnons and microwave photons in
  the quantum limit},\ }}\href {\doibase 10.1103/PhysRevLett.113.083603}
  {\bibfield  {journal} {\bibinfo  {journal} {Phys. Rev. Lett.}\ }\textbf
  {\bibinfo {volume} {113}},\ \bibinfo {pages} {083603} (\bibinfo {year}
  {2014})}\BibitemShut {NoStop}%
\bibitem [{\citenamefont {Zhang}\ \emph {et~al.}(2014)\citenamefont {Zhang},
  \citenamefont {Zou}, \citenamefont {Jiang},\ and\ \citenamefont
  {Tang}}]{yigcavity2}%
  \BibitemOpen
  \bibfield  {author} {\bibinfo {author} {\bibfnamefont {X.}~\bibnamefont
  {Zhang}}, \bibinfo {author} {\bibfnamefont {C.-L.}\ \bibnamefont {Zou}},
  \bibinfo {author} {\bibfnamefont {L.}~\bibnamefont {Jiang}}, \ and\ \bibinfo
  {author} {\bibfnamefont {H.~X.}\ \bibnamefont {Tang}},\ }\bibfield  {title}
  {\emph {\bibinfo {title} {Strongly coupled magnons and cavity microwave
  photons},\ }}\href {\doibase 10.1103/PhysRevLett.113.156401} {\bibfield
  {journal} {\bibinfo  {journal} {Phys. Rev. Lett.}\ }\textbf {\bibinfo
  {volume} {113}},\ \bibinfo {pages} {156401} (\bibinfo {year}
  {2014})}\BibitemShut {NoStop}%
\bibitem [{\citenamefont {Wang}\ \emph {et~al.}(2018)\citenamefont {Wang},
  \citenamefont {Zhang}, \citenamefont {Zhang}, \citenamefont {Li},
  \citenamefont {Hu},\ and\ \citenamefont {You}}]{kerrmagnon}%
  \BibitemOpen
  \bibfield  {author} {\bibinfo {author} {\bibfnamefont {Y.-P.}\ \bibnamefont
  {Wang}}, \bibinfo {author} {\bibfnamefont {G.-Q.}\ \bibnamefont {Zhang}},
  \bibinfo {author} {\bibfnamefont {D.}~\bibnamefont {Zhang}}, \bibinfo
  {author} {\bibfnamefont {T.-F.}\ \bibnamefont {Li}}, \bibinfo {author}
  {\bibfnamefont {C.-M.}\ \bibnamefont {Hu}}, \ and\ \bibinfo {author}
  {\bibfnamefont {J.~Q.}\ \bibnamefont {You}},\ }\bibfield  {title} {\emph
  {\bibinfo {title} {Bistability of cavity magnon polaritons},\ }}\href
  {\doibase 10.1103/PhysRevLett.120.057202} {\bibfield  {journal} {\bibinfo
  {journal} {Phys. Rev. Lett.}\ }\textbf {\bibinfo {volume} {120}},\ \bibinfo
  {pages} {057202} (\bibinfo {year} {2018})}\BibitemShut {NoStop}%
\bibitem [{\citenamefont {Shen}\ \emph {et~al.}(2021)\citenamefont {Shen},
  \citenamefont {Wang}, \citenamefont {Li}, \citenamefont {Zhu}, \citenamefont
  {Agarwal},\ and\ \citenamefont {You}}]{gatemagnon}%
  \BibitemOpen
  \bibfield  {author} {\bibinfo {author} {\bibfnamefont {R.-C.}\ \bibnamefont
  {Shen}}, \bibinfo {author} {\bibfnamefont {Y.-P.}\ \bibnamefont {Wang}},
  \bibinfo {author} {\bibfnamefont {J.}~\bibnamefont {Li}}, \bibinfo {author}
  {\bibfnamefont {S.-Y.}\ \bibnamefont {Zhu}}, \bibinfo {author} {\bibfnamefont
  {G.~S.}\ \bibnamefont {Agarwal}}, \ and\ \bibinfo {author} {\bibfnamefont
  {J.~Q.}\ \bibnamefont {You}},\ }\bibfield  {title} {\emph {\bibinfo {title}
  {Long-time memory and ternary logic gate using a multistable cavity magnonic
  system},\ }}\href {\doibase 10.1103/PhysRevLett.127.183202} {\bibfield
  {journal} {\bibinfo  {journal} {Phys. Rev. Lett.}\ }\textbf {\bibinfo
  {volume} {127}},\ \bibinfo {pages} {183202} (\bibinfo {year}
  {2021})}\BibitemShut {NoStop}%
\bibitem [{\citenamefont {Lachance-Quirion}\ \emph {et~al.}(2020)\citenamefont
  {Lachance-Quirion}, \citenamefont {Piotr~Wolski}, \citenamefont {Tabuchi},
  \citenamefont {Kono}, \citenamefont {Usami},\ and\ \citenamefont
  {Nakamura}}]{magnonqubit2}%
  \BibitemOpen
  \bibfield  {author} {\bibinfo {author} {\bibfnamefont {D.}~\bibnamefont
  {Lachance-Quirion}}, \bibinfo {author} {\bibfnamefont {S.}~\bibnamefont
  {Piotr~Wolski}}, \bibinfo {author} {\bibfnamefont {Y.}~\bibnamefont
  {Tabuchi}}, \bibinfo {author} {\bibfnamefont {S.}~\bibnamefont {Kono}},
  \bibinfo {author} {\bibfnamefont {K.}~\bibnamefont {Usami}}, \ and\ \bibinfo
  {author} {\bibfnamefont {Y.}~\bibnamefont {Nakamura}},\ }\bibfield  {title}
  {\emph {\bibinfo {title} {Entanglement-based single-shot detection of a
  single magnon with a superconducting qubit},\ }}\href {\doibase
  10.1126/science.aaz9236} {\bibfield  {journal} {\bibinfo  {journal}
  {Science}\ }\textbf {\bibinfo {volume} {367}},\ \bibinfo {pages} {425}
  (\bibinfo {year} {2020})}\BibitemShut {NoStop}%
\bibitem [{\citenamefont {Tabuchi}\ \emph {et~al.}(2015)\citenamefont
  {Tabuchi}, \citenamefont {Ishino}, \citenamefont {Noguchi}, \citenamefont
  {Ishikawa}, \citenamefont {Yamazaki}, \citenamefont {Usami},\ and\
  \citenamefont {Nakamura}}]{magnonqubit}%
  \BibitemOpen
  \bibfield  {author} {\bibinfo {author} {\bibfnamefont {Y.}~\bibnamefont
  {Tabuchi}}, \bibinfo {author} {\bibfnamefont {S.}~\bibnamefont {Ishino}},
  \bibinfo {author} {\bibfnamefont {A.}~\bibnamefont {Noguchi}}, \bibinfo
  {author} {\bibfnamefont {T.}~\bibnamefont {Ishikawa}}, \bibinfo {author}
  {\bibfnamefont {R.}~\bibnamefont {Yamazaki}}, \bibinfo {author}
  {\bibfnamefont {K.}~\bibnamefont {Usami}}, \ and\ \bibinfo {author}
  {\bibfnamefont {Y.}~\bibnamefont {Nakamura}},\ }\bibfield  {title} {\emph
  {\bibinfo {title} {Coherent coupling between a ferromagnetic magnon and a
  superconducting qubit},\ }}\href {\doibase 10.1126/science.aaa3693}
  {\bibfield  {journal} {\bibinfo  {journal} {Science}\ }\textbf {\bibinfo
  {volume} {349}},\ \bibinfo {pages} {405} (\bibinfo {year}
  {2015})}\BibitemShut {NoStop}%
\bibitem [{\citenamefont {Qi}\ and\ \citenamefont
  {Jing}(2022{\natexlab{a}})}]{magnonqubit3}%
  \BibitemOpen
  \bibfield  {author} {\bibinfo {author} {\bibfnamefont {S.-f.}\ \bibnamefont
  {Qi}}\ and\ \bibinfo {author} {\bibfnamefont {J.}~\bibnamefont {Jing}},\
  }\bibfield  {title} {\emph {\bibinfo {title} {Generation of bell and
  greenberger-horne-zeilinger states from a hybrid qubit-photon-magnon
  system},\ }}\href {\doibase 10.1103/PhysRevA.105.022624} {\bibfield
  {journal} {\bibinfo  {journal} {Phys. Rev. A}\ }\textbf {\bibinfo {volume}
  {105}},\ \bibinfo {pages} {022624} (\bibinfo {year}
  {2022}{\natexlab{a}})}\BibitemShut {NoStop}%
\bibitem [{\citenamefont {Yan}\ and\ \citenamefont
  {Jing}(2021)}]{magnonqubit4}%
  \BibitemOpen
  \bibfield  {author} {\bibinfo {author} {\bibfnamefont {J.-s.}\ \bibnamefont
  {Yan}}\ and\ \bibinfo {author} {\bibfnamefont {J.}~\bibnamefont {Jing}},\
  }\bibfield  {title} {\emph {\bibinfo {title} {External-level assisted cooling
  by measurement},\ }}\href {\doibase 10.1103/PhysRevA.104.063105} {\bibfield
  {journal} {\bibinfo  {journal} {Phys. Rev. A}\ }\textbf {\bibinfo {volume}
  {104}},\ \bibinfo {pages} {063105} (\bibinfo {year} {2021})}\BibitemShut
  {NoStop}%
\bibitem [{\citenamefont {Neuman}\ \emph {et~al.}(2020)\citenamefont {Neuman},
  \citenamefont {Wang},\ and\ \citenamefont {Narang}}]{magnonspin}%
  \BibitemOpen
  \bibfield  {author} {\bibinfo {author} {\bibfnamefont {T.~c.~v.}\
  \bibnamefont {Neuman}}, \bibinfo {author} {\bibfnamefont {D.~S.}\
  \bibnamefont {Wang}}, \ and\ \bibinfo {author} {\bibfnamefont
  {P.}~\bibnamefont {Narang}},\ }\bibfield  {title} {\emph {\bibinfo {title}
  {Nanomagnonic cavities for strong spin-magnon coupling and magnon-mediated
  spin-spin interactions},\ }}\href {\doibase 10.1103/PhysRevLett.125.247702}
  {\bibfield  {journal} {\bibinfo  {journal} {Phys. Rev. Lett.}\ }\textbf
  {\bibinfo {volume} {125}},\ \bibinfo {pages} {247702} (\bibinfo {year}
  {2020})}\BibitemShut {NoStop}%
\bibitem [{\citenamefont {Qi}\ and\ \citenamefont
  {Jing}(2021{\natexlab{a}})}]{magnonspin2}%
  \BibitemOpen
  \bibfield  {author} {\bibinfo {author} {\bibfnamefont {S.-f.}\ \bibnamefont
  {Qi}}\ and\ \bibinfo {author} {\bibfnamefont {J.}~\bibnamefont {Jing}},\
  }\bibfield  {title} {\emph {\bibinfo {title} {Magnon-mediated quantum battery
  under systematic errors},\ }}\href {\doibase 10.1103/PhysRevA.104.032606}
  {\bibfield  {journal} {\bibinfo  {journal} {Phys. Rev. A}\ }\textbf {\bibinfo
  {volume} {104}},\ \bibinfo {pages} {032606} (\bibinfo {year}
  {2021}{\natexlab{a}})}\BibitemShut {NoStop}%
\bibitem [{\citenamefont {Zhang}\ \emph {et~al.}(2016)\citenamefont {Zhang},
  \citenamefont {Zou}, \citenamefont {Jiang},\ and\ \citenamefont
  {Tang}}]{magnoncavity}%
  \BibitemOpen
  \bibfield  {author} {\bibinfo {author} {\bibfnamefont {X.}~\bibnamefont
  {Zhang}}, \bibinfo {author} {\bibfnamefont {C.-L.}\ \bibnamefont {Zou}},
  \bibinfo {author} {\bibfnamefont {L.}~\bibnamefont {Jiang}}, \ and\ \bibinfo
  {author} {\bibfnamefont {H.}~\bibnamefont {Tang}},\ }\bibfield  {title}
  {\emph {\bibinfo {title} {Cavity magnonmechanics},\ }}\href {\doibase
  10.1126/sciadv.1501286} {\bibfield  {journal} {\bibinfo  {journal} {Sci.
  Adv.}\ }\textbf {\bibinfo {volume} {2}},\ \bibinfo {pages} {e1501286}
  (\bibinfo {year} {2016})}\BibitemShut {NoStop}%
\bibitem [{\citenamefont {Li}\ \emph {et~al.}(2018{\natexlab{a}})\citenamefont
  {Li}, \citenamefont {Zhu},\ and\ \citenamefont {Agarwal}}]{mppentang}%
  \BibitemOpen
  \bibfield  {author} {\bibinfo {author} {\bibfnamefont {J.}~\bibnamefont
  {Li}}, \bibinfo {author} {\bibfnamefont {S.-Y.}\ \bibnamefont {Zhu}}, \ and\
  \bibinfo {author} {\bibfnamefont {G.~S.}\ \bibnamefont {Agarwal}},\
  }\bibfield  {title} {\emph {\bibinfo {title} {Magnon-photon-phonon
  entanglement in cavity magnomechanics},\ }}\href {\doibase
  10.1103/PhysRevLett.121.203601} {\bibfield  {journal} {\bibinfo  {journal}
  {Phys. Rev. Lett.}\ }\textbf {\bibinfo {volume} {121}},\ \bibinfo {pages}
  {203601} (\bibinfo {year} {2018}{\natexlab{a}})}\BibitemShut {NoStop}%
\bibitem [{\citenamefont {Qi}\ and\ \citenamefont
  {Jing}(2021{\natexlab{b}})}]{intermagnon}%
  \BibitemOpen
  \bibfield  {author} {\bibinfo {author} {\bibfnamefont {S.-f.}\ \bibnamefont
  {Qi}}\ and\ \bibinfo {author} {\bibfnamefont {J.}~\bibnamefont {Jing}},\
  }\bibfield  {title} {\emph {\bibinfo {title} {Magnon-assisted photon-phonon
  conversion in the presence of structured environments},\ }}\href {\doibase
  10.1103/PhysRevA.103.043704} {\bibfield  {journal} {\bibinfo  {journal}
  {Phys. Rev. A}\ }\textbf {\bibinfo {volume} {103}},\ \bibinfo {pages}
  {043704} (\bibinfo {year} {2021}{\natexlab{b}})}\BibitemShut {NoStop}%
\bibitem [{\citenamefont {Li}\ \emph {et~al.}(2021)\citenamefont {Li},
  \citenamefont {Wang}, \citenamefont {Wu}, \citenamefont {Zhu},\ and\
  \citenamefont {You}}]{quantumnetwork}%
  \BibitemOpen
  \bibfield  {author} {\bibinfo {author} {\bibfnamefont {J.}~\bibnamefont
  {Li}}, \bibinfo {author} {\bibfnamefont {Y.-P.}\ \bibnamefont {Wang}},
  \bibinfo {author} {\bibfnamefont {W.-J.}\ \bibnamefont {Wu}}, \bibinfo
  {author} {\bibfnamefont {S.-Y.}\ \bibnamefont {Zhu}}, \ and\ \bibinfo
  {author} {\bibfnamefont {J.}~\bibnamefont {You}},\ }\bibfield  {title} {\emph
  {\bibinfo {title} {Quantum network with magnonic and mechanical nodes},\
  }}\href {\doibase 10.1103/PRXQuantum.2.040344} {\bibfield  {journal}
  {\bibinfo  {journal} {PRX Quant.}\ }\textbf {\bibinfo {volume} {2}},\
  \bibinfo {pages} {040344} (\bibinfo {year} {2021})}\BibitemShut {NoStop}%
\bibitem [{\citenamefont {Hatanaka}\ \emph {et~al.}(2022)\citenamefont
  {Hatanaka}, \citenamefont {Asano}, \citenamefont {Okamoto}, \citenamefont
  {Kunihashi}, \citenamefont {Sanada},\ and\ \citenamefont
  {Yamaguchi}}]{magnonphonononchip}%
  \BibitemOpen
  \bibfield  {author} {\bibinfo {author} {\bibfnamefont {D.}~\bibnamefont
  {Hatanaka}}, \bibinfo {author} {\bibfnamefont {M.}~\bibnamefont {Asano}},
  \bibinfo {author} {\bibfnamefont {H.}~\bibnamefont {Okamoto}}, \bibinfo
  {author} {\bibfnamefont {Y.}~\bibnamefont {Kunihashi}}, \bibinfo {author}
  {\bibfnamefont {H.}~\bibnamefont {Sanada}}, \ and\ \bibinfo {author}
  {\bibfnamefont {H.}~\bibnamefont {Yamaguchi}},\ }\bibfield  {title} {\emph
  {\bibinfo {title} {On-chip coherent transduction between magnons and acoustic
  phonons in cavity magnomechanics},\ }}\href {\doibase
  10.1103/PhysRevApplied.17.034024} {\bibfield  {journal} {\bibinfo  {journal}
  {Phys. Rev. Appl.}\ }\textbf {\bibinfo {volume} {17}},\ \bibinfo {pages}
  {034024} (\bibinfo {year} {2022})}\BibitemShut {NoStop}%
\bibitem [{\citenamefont {Potts}\ \emph {et~al.}(2021)\citenamefont {Potts},
  \citenamefont {Varga}, \citenamefont {Bittencourt}, \citenamefont
  {Kusminskiy},\ and\ \citenamefont {Davis}}]{dynamicalmagnon}%
  \BibitemOpen
  \bibfield  {author} {\bibinfo {author} {\bibfnamefont {C.~A.}\ \bibnamefont
  {Potts}}, \bibinfo {author} {\bibfnamefont {E.}~\bibnamefont {Varga}},
  \bibinfo {author} {\bibfnamefont {V.~A. S.~V.}\ \bibnamefont {Bittencourt}},
  \bibinfo {author} {\bibfnamefont {S.~V.}\ \bibnamefont {Kusminskiy}}, \ and\
  \bibinfo {author} {\bibfnamefont {J.~P.}\ \bibnamefont {Davis}},\ }\bibfield
  {title} {\emph {\bibinfo {title} {Dynamical backaction magnomechanics},\
  }}\href {\doibase 10.1103/PhysRevX.11.031053} {\bibfield  {journal} {\bibinfo
   {journal} {Phys. Rev. X}\ }\textbf {\bibinfo {volume} {11}},\ \bibinfo
  {pages} {031053} (\bibinfo {year} {2021})}\BibitemShut {NoStop}%
\bibitem [{\citenamefont {Shen}\ \emph {et~al.}(2022)\citenamefont {Shen},
  \citenamefont {Li}, \citenamefont {Fan}, \citenamefont {Wang},\ and\
  \citenamefont {You}}]{kerrmagnomech}%
  \BibitemOpen
  \bibfield  {author} {\bibinfo {author} {\bibfnamefont {R.-C.}\ \bibnamefont
  {Shen}}, \bibinfo {author} {\bibfnamefont {J.}~\bibnamefont {Li}}, \bibinfo
  {author} {\bibfnamefont {Z.-y.}\ \bibnamefont {Fan}}, \bibinfo {author}
  {\bibfnamefont {Y.-P.}\ \bibnamefont {Wang}}, \ and\ \bibinfo {author}
  {\bibfnamefont {J.}~\bibnamefont {You}},\ }\bibfield  {title} {\emph
  {\bibinfo {title} {Mechanical bistability in kerr-modeified cavity
  magnomechanics},\ }}\href@noop {} {\bibfield  {journal} {\bibinfo  {journal}
  {arxiv: 2206. 14588}\ } (\bibinfo {year} {2022})}\BibitemShut {NoStop}%
\bibitem [{\citenamefont {Ladd}\ \emph {et~al.}(2010)\citenamefont {Ladd},
  \citenamefont {Jelezko}, \citenamefont {Laflamme}, \citenamefont {Nakamura},
  \citenamefont {Monroe},\ and\ \citenamefont {O'Brien}}]{quantumcomputing}%
  \BibitemOpen
  \bibfield  {author} {\bibinfo {author} {\bibfnamefont {T.~D.}\ \bibnamefont
  {Ladd}}, \bibinfo {author} {\bibfnamefont {F.}~\bibnamefont {Jelezko}},
  \bibinfo {author} {\bibfnamefont {R.}~\bibnamefont {Laflamme}}, \bibinfo
  {author} {\bibfnamefont {Y.}~\bibnamefont {Nakamura}}, \bibinfo {author}
  {\bibfnamefont {C.}~\bibnamefont {Monroe}}, \ and\ \bibinfo {author}
  {\bibfnamefont {J.~L.}\ \bibnamefont {O'Brien}},\ }\bibfield  {title} {\emph
  {\bibinfo {title} {Quantum computers},\ }}\href {\doibase
  https://doi.org/10.1038/nature08812} {\bibfield  {journal} {\bibinfo
  {journal} {Nature (London)}\ }\textbf {\bibinfo {volume} {464}},\ \bibinfo
  {pages} {45} (\bibinfo {year} {2010})}\BibitemShut {NoStop}%
\bibitem [{\citenamefont {Reiserer}\ and\ \citenamefont
  {Rempe}(2015)}]{quantumcommunication}%
  \BibitemOpen
  \bibfield  {author} {\bibinfo {author} {\bibfnamefont {A.}~\bibnamefont
  {Reiserer}}\ and\ \bibinfo {author} {\bibfnamefont {G.}~\bibnamefont
  {Rempe}},\ }\bibfield  {title} {\emph {\bibinfo {title} {Cavity-based quantum
  networks with single atoms and optical photons},\ }}\href {\doibase
  10.1103/RevModPhys.87.1379} {\bibfield  {journal} {\bibinfo  {journal} {Rev.
  Mod. Phys.}\ }\textbf {\bibinfo {volume} {87}},\ \bibinfo {pages} {1379}
  (\bibinfo {year} {2015})}\BibitemShut {NoStop}%
\bibitem [{\citenamefont {Degen}\ \emph {et~al.}(2017)\citenamefont {Degen},
  \citenamefont {Reinhard},\ and\ \citenamefont {Cappellaro}}]{quantumsense}%
  \BibitemOpen
  \bibfield  {author} {\bibinfo {author} {\bibfnamefont {C.~L.}\ \bibnamefont
  {Degen}}, \bibinfo {author} {\bibfnamefont {F.}~\bibnamefont {Reinhard}}, \
  and\ \bibinfo {author} {\bibfnamefont {P.}~\bibnamefont {Cappellaro}},\
  }\bibfield  {title} {\emph {\bibinfo {title} {Quantum sensing},\ }}\href
  {\doibase 10.1103/RevModPhys.89.035002} {\bibfield  {journal} {\bibinfo
  {journal} {Rev. Mod. Phys.}\ }\textbf {\bibinfo {volume} {89}},\ \bibinfo
  {pages} {035002} (\bibinfo {year} {2017})}\BibitemShut {NoStop}%
\bibitem [{\citenamefont {Xu}\ \emph {et~al.}(2020)\citenamefont {Xu},
  \citenamefont {Zhong}, \citenamefont {Han}, \citenamefont {Jin},
  \citenamefont {Jiang},\ and\ \citenamefont {Zhang}}]{Floquetmagnon}%
  \BibitemOpen
  \bibfield  {author} {\bibinfo {author} {\bibfnamefont {J.}~\bibnamefont
  {Xu}}, \bibinfo {author} {\bibfnamefont {C.}~\bibnamefont {Zhong}}, \bibinfo
  {author} {\bibfnamefont {X.}~\bibnamefont {Han}}, \bibinfo {author}
  {\bibfnamefont {D.}~\bibnamefont {Jin}}, \bibinfo {author} {\bibfnamefont
  {L.}~\bibnamefont {Jiang}}, \ and\ \bibinfo {author} {\bibfnamefont
  {X.}~\bibnamefont {Zhang}},\ }\bibfield  {title} {\emph {\bibinfo {title}
  {Floquet cavity electromagnonics},\ }}\href {\doibase
  10.1103/PhysRevLett.125.237201} {\bibfield  {journal} {\bibinfo  {journal}
  {Phys. Rev. Lett.}\ }\textbf {\bibinfo {volume} {125}},\ \bibinfo {pages}
  {237201} (\bibinfo {year} {2020})}\BibitemShut {NoStop}%
\bibitem [{\citenamefont {Lodahl}\ \emph {et~al.}(2017)\citenamefont {Lodahl},
  \citenamefont {Mahmoodian}, \citenamefont {Stobbe}, \citenamefont
  {Rauschenbeutel}, \citenamefont {Schneeweiss}, \citenamefont {Volz},
  \citenamefont {Pichler},\ and\ \citenamefont {Zoller}}]{chiral}%
  \BibitemOpen
  \bibfield  {author} {\bibinfo {author} {\bibfnamefont {P.}~\bibnamefont
  {Lodahl}}, \bibinfo {author} {\bibfnamefont {S.}~\bibnamefont {Mahmoodian}},
  \bibinfo {author} {\bibfnamefont {S.}~\bibnamefont {Stobbe}}, \bibinfo
  {author} {\bibfnamefont {A.}~\bibnamefont {Rauschenbeutel}}, \bibinfo
  {author} {\bibfnamefont {P.}~\bibnamefont {Schneeweiss}}, \bibinfo {author}
  {\bibfnamefont {J.}~\bibnamefont {Volz}}, \bibinfo {author} {\bibfnamefont
  {H.}~\bibnamefont {Pichler}}, \ and\ \bibinfo {author} {\bibfnamefont
  {P.}~\bibnamefont {Zoller}},\ }\bibfield  {title} {\emph {\bibinfo {title}
  {Chiral quantum optics},\ }}\href {\doibase 10.1038/nature21037} {\bibfield
  {journal} {\bibinfo  {journal} {Nature}\ }\textbf {\bibinfo {volume} {541}},\
  \bibinfo {pages} {473} (\bibinfo {year} {2017})}\BibitemShut {NoStop}%
\bibitem [{\citenamefont {Zhu}\ \emph {et~al.}(2018)\citenamefont {Zhu},
  \citenamefont {Yi}, \citenamefont {Li}, \citenamefont {Xiao}, \citenamefont
  {Zhang}, \citenamefont {Yang}, \citenamefont {Kaindl}, \citenamefont {Li},
  \citenamefont {Wang},\ and\ \citenamefont {Zhang}}]{chiral2}%
  \BibitemOpen
  \bibfield  {author} {\bibinfo {author} {\bibfnamefont {H.}~\bibnamefont
  {Zhu}}, \bibinfo {author} {\bibfnamefont {J.}~\bibnamefont {Yi}}, \bibinfo
  {author} {\bibfnamefont {M.-Y.}\ \bibnamefont {Li}}, \bibinfo {author}
  {\bibfnamefont {J.}~\bibnamefont {Xiao}}, \bibinfo {author} {\bibfnamefont
  {L.}~\bibnamefont {Zhang}}, \bibinfo {author} {\bibfnamefont {C.-W.}\
  \bibnamefont {Yang}}, \bibinfo {author} {\bibfnamefont {R.~A.}\ \bibnamefont
  {Kaindl}}, \bibinfo {author} {\bibfnamefont {L.-J.}\ \bibnamefont {Li}},
  \bibinfo {author} {\bibfnamefont {Y.}~\bibnamefont {Wang}}, \ and\ \bibinfo
  {author} {\bibfnamefont {X.}~\bibnamefont {Zhang}},\ }\bibfield  {title}
  {\emph {\bibinfo {title} {Observation of chiral phonons},\ }}\href {\doibase
  10.1126/science.aar2711} {\bibfield  {journal} {\bibinfo  {journal}
  {Science}\ }\textbf {\bibinfo {volume} {359}},\ \bibinfo {pages} {579}
  (\bibinfo {year} {2018})}\BibitemShut {NoStop}%
\bibitem [{\citenamefont {S\o{}rensen}\ \emph {et~al.}(2005)\citenamefont
  {S\o{}rensen}, \citenamefont {Demler},\ and\ \citenamefont
  {Lukin}}]{fraction}%
  \BibitemOpen
  \bibfield  {author} {\bibinfo {author} {\bibfnamefont {A.~S.}\ \bibnamefont
  {S\o{}rensen}}, \bibinfo {author} {\bibfnamefont {E.}~\bibnamefont {Demler}},
  \ and\ \bibinfo {author} {\bibfnamefont {M.~D.}\ \bibnamefont {Lukin}},\
  }\bibfield  {title} {\emph {\bibinfo {title} {Fractional quantum hall states
  of atoms in optical lattices},\ }}\href {\doibase
  10.1103/PhysRevLett.94.086803} {\bibfield  {journal} {\bibinfo  {journal}
  {Phys. Rev. Lett.}\ }\textbf {\bibinfo {volume} {94}},\ \bibinfo {pages}
  {086803} (\bibinfo {year} {2005})}\BibitemShut {NoStop}%
\bibitem [{\citenamefont {Liu}\ \emph {et~al.}(2020)\citenamefont {Liu},
  \citenamefont {Feng}, \citenamefont {Ren}, \citenamefont {Wang},\ and\
  \citenamefont {Wang}}]{chiralspin2}%
  \BibitemOpen
  \bibfield  {author} {\bibinfo {author} {\bibfnamefont {W.}~\bibnamefont
  {Liu}}, \bibinfo {author} {\bibfnamefont {W.}~\bibnamefont {Feng}}, \bibinfo
  {author} {\bibfnamefont {W.}~\bibnamefont {Ren}}, \bibinfo {author}
  {\bibfnamefont {D.-W.}\ \bibnamefont {Wang}}, \ and\ \bibinfo {author}
  {\bibfnamefont {H.}~\bibnamefont {Wang}},\ }\bibfield  {title} {\emph
  {\bibinfo {title} {Synthesizing three-body interaction of spin chirality with
  superconducting qubits},\ }}\href {\doibase 10.1063/1.5140884} {\bibfield
  {journal} {\bibinfo  {journal} {Appl. Phys. Lett.}\ }\textbf {\bibinfo
  {volume} {116}},\ \bibinfo {pages} {114001} (\bibinfo {year}
  {2020})}\BibitemShut {NoStop}%
\bibitem [{\citenamefont {Cai}\ and\ \citenamefont {Wang}(2021)}]{topolight}%
  \BibitemOpen
  \bibfield  {author} {\bibinfo {author} {\bibfnamefont {H.}~\bibnamefont
  {Cai}}\ and\ \bibinfo {author} {\bibfnamefont {D.-W.}\ \bibnamefont {Wang}},\
  }\bibfield  {title} {\emph {\bibinfo {title} {Topological phases of quantized
  light},\ }}\href {\doibase 10.1093/nsr/nwaa196} {\bibfield  {journal}
  {\bibinfo  {journal} {Nat. Sci. Rev.}\ }\textbf {\bibinfo {volume} {8}},\
  \bibinfo {pages} {nwaa196} (\bibinfo {year} {2021})}\BibitemShut {NoStop}%
\bibitem [{\citenamefont {Shirley}(1965)}]{Floquet}%
  \BibitemOpen
  \bibfield  {author} {\bibinfo {author} {\bibfnamefont {J.~H.}\ \bibnamefont
  {Shirley}},\ }\bibfield  {title} {\emph {\bibinfo {title} {Solution of the
  schr\"odinger equation with a hamiltonian periodic in time},\ }}\href
  {\doibase 10.1103/PhysRev.138.B979} {\bibfield  {journal} {\bibinfo
  {journal} {Phys. Rev.}\ }\textbf {\bibinfo {volume} {138}},\ \bibinfo {pages}
  {B979} (\bibinfo {year} {1965})}\BibitemShut {NoStop}%
\bibitem [{\citenamefont {Goldman}\ and\ \citenamefont
  {Dalibard}(2014)}]{Floquet1}%
  \BibitemOpen
  \bibfield  {author} {\bibinfo {author} {\bibfnamefont {N.}~\bibnamefont
  {Goldman}}\ and\ \bibinfo {author} {\bibfnamefont {J.}~\bibnamefont
  {Dalibard}},\ }\bibfield  {title} {\emph {\bibinfo {title} {Periodically
  driven quantum systems: Effective hamiltonians and engineered gauge fields},\
  }}\href {\doibase 10.1103/PhysRevX.4.031027} {\bibfield  {journal} {\bibinfo
  {journal} {Phys. Rev. X}\ }\textbf {\bibinfo {volume} {4}},\ \bibinfo {pages}
  {031027} (\bibinfo {year} {2014})}\BibitemShut {NoStop}%
\bibitem [{\citenamefont {Bukov}\ \emph {et~al.}(2015)\citenamefont {Bukov},
  \citenamefont {D'Alessio},\ and\ \citenamefont
  {Polkovnikov}}]{Floquettheory}%
  \BibitemOpen
  \bibfield  {author} {\bibinfo {author} {\bibfnamefont {M.}~\bibnamefont
  {Bukov}}, \bibinfo {author} {\bibfnamefont {L.}~\bibnamefont {D'Alessio}}, \
  and\ \bibinfo {author} {\bibfnamefont {A.}~\bibnamefont {Polkovnikov}},\
  }\bibfield  {title} {\emph {\bibinfo {title} {Universal high-frequency
  behavior of periodically driven system: from dynamical stabilization to
  floquet engineering},\ }}\href {\doibase 10.1080/00018732.2015.1055918}
  {\bibfield  {journal} {\bibinfo  {journal} {Adv. in Phys.}\ }\textbf
  {\bibinfo {volume} {64}},\ \bibinfo {pages} {139} (\bibinfo {year}
  {2015})}\BibitemShut {NoStop}%
\bibitem [{\citenamefont {Petiziol}\ \emph {et~al.}(2021)\citenamefont
  {Petiziol}, \citenamefont {Sameti}, \citenamefont {Carretta}, \citenamefont
  {Wimberger},\ and\ \citenamefont {Mintert}}]{Floquet3}%
  \BibitemOpen
  \bibfield  {author} {\bibinfo {author} {\bibfnamefont {F.}~\bibnamefont
  {Petiziol}}, \bibinfo {author} {\bibfnamefont {M.}~\bibnamefont {Sameti}},
  \bibinfo {author} {\bibfnamefont {S.}~\bibnamefont {Carretta}}, \bibinfo
  {author} {\bibfnamefont {S.}~\bibnamefont {Wimberger}}, \ and\ \bibinfo
  {author} {\bibfnamefont {F.}~\bibnamefont {Mintert}},\ }\bibfield  {title}
  {\emph {\bibinfo {title} {Quantum simulation of three-body interactions in
  weakly driven quantum systems},\ }}\href {\doibase
  10.1103/PhysRevLett.126.250504} {\bibfield  {journal} {\bibinfo  {journal}
  {Phys. Rev. Lett.}\ }\textbf {\bibinfo {volume} {126}},\ \bibinfo {pages}
  {250504} (\bibinfo {year} {2021})}\BibitemShut {NoStop}%
\bibitem [{\citenamefont {Shao}\ \emph {et~al.}(2017)\citenamefont {Shao},
  \citenamefont {Wu},\ and\ \citenamefont {Feng}}]{James}%
  \BibitemOpen
  \bibfield  {author} {\bibinfo {author} {\bibfnamefont {W.}~\bibnamefont
  {Shao}}, \bibinfo {author} {\bibfnamefont {C.}~\bibnamefont {Wu}}, \ and\
  \bibinfo {author} {\bibfnamefont {X.-L.}\ \bibnamefont {Feng}},\ }\bibfield
  {title} {\emph {\bibinfo {title} {Generalized james' effective hamiltonian
  method},\ }}\href {\doibase 10.1103/PhysRevA.95.032124} {\bibfield  {journal}
  {\bibinfo  {journal} {Phys. Rev. A}\ }\textbf {\bibinfo {volume} {95}},\
  \bibinfo {pages} {032124} (\bibinfo {year} {2017})}\BibitemShut {NoStop}%
\bibitem [{\citenamefont {Wang}\ \emph
  {et~al.}(2016{\natexlab{a}})\citenamefont {Wang}, \citenamefont {Cai},
  \citenamefont {Liu},\ and\ \citenamefont {Scully}}]{Floquetnoon}%
  \BibitemOpen
  \bibfield  {author} {\bibinfo {author} {\bibfnamefont {D.-W.}\ \bibnamefont
  {Wang}}, \bibinfo {author} {\bibfnamefont {H.}~\bibnamefont {Cai}}, \bibinfo
  {author} {\bibfnamefont {R.-B.}\ \bibnamefont {Liu}}, \ and\ \bibinfo
  {author} {\bibfnamefont {M.~O.}\ \bibnamefont {Scully}},\ }\bibfield  {title}
  {\emph {\bibinfo {title} {Mesoscopic superposition states generated by
  synthetic spin-orbit interaction in fock-state lattices},\ }}\href {\doibase
  10.1103/PhysRevLett.116.220502} {\bibfield  {journal} {\bibinfo  {journal}
  {Phys. Rev. Lett.}\ }\textbf {\bibinfo {volume} {116}},\ \bibinfo {pages}
  {220502} (\bibinfo {year} {2016}{\natexlab{a}})}\BibitemShut {NoStop}%
\bibitem [{\citenamefont {Li}\ \emph {et~al.}(2019)\citenamefont {Li},
  \citenamefont {Cai}, \citenamefont {Xu}, \citenamefont {Yakovlev},
  \citenamefont {Yang},\ and\ \citenamefont {Wang}}]{Floquet2}%
  \BibitemOpen
  \bibfield  {author} {\bibinfo {author} {\bibfnamefont {H.}~\bibnamefont
  {Li}}, \bibinfo {author} {\bibfnamefont {H.}~\bibnamefont {Cai}}, \bibinfo
  {author} {\bibfnamefont {J.}~\bibnamefont {Xu}}, \bibinfo {author}
  {\bibfnamefont {V.~V.}\ \bibnamefont {Yakovlev}}, \bibinfo {author}
  {\bibfnamefont {Y.}~\bibnamefont {Yang}}, \ and\ \bibinfo {author}
  {\bibfnamefont {D.-W.}\ \bibnamefont {Wang}},\ }\bibfield  {title} {\emph
  {\bibinfo {title} {Quantum photonic transistor controlled by an atom in a
  floquet cavity-qed system},\ }}\href {\doibase 10.1364/OE.27.006946}
  {\bibfield  {journal} {\bibinfo  {journal} {Opt. Express}\ }\textbf {\bibinfo
  {volume} {27}},\ \bibinfo {pages} {6946} (\bibinfo {year}
  {2019})}\BibitemShut {NoStop}%
\bibitem [{\citenamefont {Wu}\ \emph {et~al.}(2018)\citenamefont {Wu},
  \citenamefont {Yang}, \citenamefont {Gong}, \citenamefont {Zheng},
  \citenamefont {Deng}, \citenamefont {Yan}, \citenamefont {Zhao},
  \citenamefont {Huang}, \citenamefont {Castellano}, \citenamefont {Munro},
  \citenamefont {Nemoto}, \citenamefont {Zheng}, \citenamefont {Sun},
  \citenamefont {Liu}, \citenamefont {Zhu},\ and\ \citenamefont {Lu}}]{switch}%
  \BibitemOpen
  \bibfield  {author} {\bibinfo {author} {\bibfnamefont {Y.}~\bibnamefont
  {Wu}}, \bibinfo {author} {\bibfnamefont {L.-P.}\ \bibnamefont {Yang}},
  \bibinfo {author} {\bibfnamefont {M.}~\bibnamefont {Gong}}, \bibinfo {author}
  {\bibfnamefont {Y.}~\bibnamefont {Zheng}}, \bibinfo {author} {\bibfnamefont
  {H.}~\bibnamefont {Deng}}, \bibinfo {author} {\bibfnamefont {Z.}~\bibnamefont
  {Yan}}, \bibinfo {author} {\bibfnamefont {Y.}~\bibnamefont {Zhao}}, \bibinfo
  {author} {\bibfnamefont {K.}~\bibnamefont {Huang}}, \bibinfo {author}
  {\bibfnamefont {A.~D.}\ \bibnamefont {Castellano}}, \bibinfo {author}
  {\bibfnamefont {W.~J.}\ \bibnamefont {Munro}}, \bibinfo {author}
  {\bibfnamefont {K.}~\bibnamefont {Nemoto}}, \bibinfo {author} {\bibfnamefont
  {D.-N.}\ \bibnamefont {Zheng}}, \bibinfo {author} {\bibfnamefont
  {C.}~\bibnamefont {Sun}}, \bibinfo {author} {\bibfnamefont {Y.-x.}\
  \bibnamefont {Liu}}, \bibinfo {author} {\bibfnamefont {X.}~\bibnamefont
  {Zhu}}, \ and\ \bibinfo {author} {\bibfnamefont {L.}~\bibnamefont {Lu}},\
  }\bibfield  {title} {\emph {\bibinfo {title} {An efficient and compact switch
  for quantum circuits},\ }}\href {\doibase 10.1038/s41534-018-0099-6}
  {\bibfield  {journal} {\bibinfo  {journal} {npj. Quantum Inf.}\ }\textbf
  {\bibinfo {volume} {4}},\ \bibinfo {pages} {50} (\bibinfo {year}
  {2018})}\BibitemShut {NoStop}%
\bibitem [{\citenamefont {Roushan1}\ \emph {et~al.}(2017)\citenamefont
  {Roushan1}, \citenamefont {Neill}, \citenamefont {Megrant}, \citenamefont
  {Chen}, \citenamefont {Babbush}, \citenamefont {Barends}, \citenamefont
  {Campbell}, \citenamefont {Chen}, \citenamefont {Chiaro}, \citenamefont
  {Dunsworth}, \citenamefont {Fowler}, \citenamefont {Jeffrey}, \citenamefont
  {Kelly}, \citenamefont {Lucero}, \citenamefont {Mutus}, \citenamefont
  {\'OMalley}, \citenamefont {Neeley}, \citenamefont {Quintana}, \citenamefont
  {Sank}, \citenamefont {Vainsencher}, \citenamefont {Wenner}, \citenamefont
  {White}, \citenamefont {Kapit}, \citenamefont {Neven},\ and\ \citenamefont
  {Martinis}}]{chiralcurrent}%
  \BibitemOpen
  \bibfield  {author} {\bibinfo {author} {\bibfnamefont {P.}~\bibnamefont
  {Roushan1}}, \bibinfo {author} {\bibfnamefont {C.}~\bibnamefont {Neill}},
  \bibinfo {author} {\bibfnamefont {A.}~\bibnamefont {Megrant}}, \bibinfo
  {author} {\bibfnamefont {Y.}~\bibnamefont {Chen}}, \bibinfo {author}
  {\bibfnamefont {R.}~\bibnamefont {Babbush}}, \bibinfo {author} {\bibfnamefont
  {R.}~\bibnamefont {Barends}}, \bibinfo {author} {\bibfnamefont
  {B.}~\bibnamefont {Campbell}}, \bibinfo {author} {\bibfnamefont
  {Z.}~\bibnamefont {Chen}}, \bibinfo {author} {\bibfnamefont {B.}~\bibnamefont
  {Chiaro}}, \bibinfo {author} {\bibfnamefont {A.}~\bibnamefont {Dunsworth}},
  \bibinfo {author} {\bibfnamefont {A.}~\bibnamefont {Fowler}}, \bibinfo
  {author} {\bibfnamefont {E.}~\bibnamefont {Jeffrey}}, \bibinfo {author}
  {\bibfnamefont {J.}~\bibnamefont {Kelly}}, \bibinfo {author} {\bibfnamefont
  {E.}~\bibnamefont {Lucero}}, \bibinfo {author} {\bibfnamefont
  {J.}~\bibnamefont {Mutus}}, \bibinfo {author} {\bibfnamefont
  {P.}~\bibnamefont {\'OMalley}}, \bibinfo {author} {\bibfnamefont
  {M.}~\bibnamefont {Neeley}}, \bibinfo {author} {\bibfnamefont
  {C.}~\bibnamefont {Quintana}}, \bibinfo {author} {\bibfnamefont
  {D.}~\bibnamefont {Sank}}, \bibinfo {author} {\bibfnamefont {A.}~\bibnamefont
  {Vainsencher}}, \bibinfo {author} {\bibfnamefont {J.}~\bibnamefont {Wenner}},
  \bibinfo {author} {\bibfnamefont {T.}~\bibnamefont {White}}, \bibinfo
  {author} {\bibfnamefont {E.}~\bibnamefont {Kapit}}, \bibinfo {author}
  {\bibfnamefont {H.}~\bibnamefont {Neven}}, \ and\ \bibinfo {author}
  {\bibfnamefont {J.}~\bibnamefont {Martinis}},\ }\bibfield  {title} {\emph
  {\bibinfo {title} {Chiral ground-state currents of interacting photons in a
  synthetic magnetic field},\ }}\href {\doibase 10.1038/NPHYS3930} {\bibfield
  {journal} {\bibinfo  {journal} {Nat. Phys.}\ }\textbf {\bibinfo {volume}
  {13}},\ \bibinfo {pages} {146} (\bibinfo {year} {2017})}\BibitemShut
  {NoStop}%
\bibitem [{\citenamefont {Kyriienko}\ and\ \citenamefont
  {S\o{}rensen}(2018)}]{Floquetsimu}%
  \BibitemOpen
  \bibfield  {author} {\bibinfo {author} {\bibfnamefont {O.}~\bibnamefont
  {Kyriienko}}\ and\ \bibinfo {author} {\bibfnamefont {A.~S.}\ \bibnamefont
  {S\o{}rensen}},\ }\bibfield  {title} {\emph {\bibinfo {title} {Floquet
  quantum simulation with superconducting qubits},\ }}\href {\doibase
  10.1103/PhysRevApplied.9.064029} {\bibfield  {journal} {\bibinfo  {journal}
  {Phys. Rev. Appl.}\ }\textbf {\bibinfo {volume} {9}},\ \bibinfo {pages}
  {064029} (\bibinfo {year} {2018})}\BibitemShut {NoStop}%
\bibitem [{\citenamefont {Wang}\ \emph {et~al.}(2019)\citenamefont {Wang},
  \citenamefont {Song}, \citenamefont {Feng},\ and\ \citenamefont
  {et~al.}}]{chiralspin}%
  \BibitemOpen
  \bibfield  {author} {\bibinfo {author} {\bibfnamefont {D.-W.}\ \bibnamefont
  {Wang}}, \bibinfo {author} {\bibfnamefont {C.}~\bibnamefont {Song}}, \bibinfo
  {author} {\bibfnamefont {W.}~\bibnamefont {Feng}}, \ and\ \bibinfo {author}
  {\bibnamefont {et~al.}},\ }\bibfield  {title} {\emph {\bibinfo {title}
  {Synthesis of antisymmetric spin exchange interaction and chiral spin
  clusters in superconducting circuits},\ }}\href {\doibase
  10.1038/s41567-018-0400-9} {\bibfield  {journal} {\bibinfo  {journal} {Nat.
  Phys.}\ }\textbf {\bibinfo {volume} {15}},\ \bibinfo {pages} {382} (\bibinfo
  {year} {2019})}\BibitemShut {NoStop}%
\bibitem [{\citenamefont {Li}\ \emph {et~al.}(2018{\natexlab{b}})\citenamefont
  {Li}, \citenamefont {Ma}, \citenamefont {Han}, \citenamefont {Chen},
  \citenamefont {Xu}, \citenamefont {Cai}, \citenamefont {Wang}, \citenamefont
  {Song}, \citenamefont {Xue}, \citenamefont {Yin},\ and\ \citenamefont
  {Sun}}]{chiralspin3}%
  \BibitemOpen
  \bibfield  {author} {\bibinfo {author} {\bibfnamefont {X.}~\bibnamefont
  {Li}}, \bibinfo {author} {\bibfnamefont {Y.}~\bibnamefont {Ma}}, \bibinfo
  {author} {\bibfnamefont {J.}~\bibnamefont {Han}}, \bibinfo {author}
  {\bibfnamefont {T.}~\bibnamefont {Chen}}, \bibinfo {author} {\bibfnamefont
  {Y.}~\bibnamefont {Xu}}, \bibinfo {author} {\bibfnamefont {W.}~\bibnamefont
  {Cai}}, \bibinfo {author} {\bibfnamefont {H.}~\bibnamefont {Wang}}, \bibinfo
  {author} {\bibfnamefont {Y.}~\bibnamefont {Song}}, \bibinfo {author}
  {\bibfnamefont {Z.-Y.}\ \bibnamefont {Xue}}, \bibinfo {author} {\bibfnamefont
  {Z.-q.}\ \bibnamefont {Yin}}, \ and\ \bibinfo {author} {\bibfnamefont
  {L.}~\bibnamefont {Sun}},\ }\bibfield  {title} {\emph {\bibinfo {title}
  {Perfect quantum state transfer in a superconducting qubit chain with
  parametrically tunable couplings},\ }}\href {\doibase
  10.1103/PhysRevApplied.10.054009} {\bibfield  {journal} {\bibinfo  {journal}
  {Phys. Rev. Appl.}\ }\textbf {\bibinfo {volume} {10}},\ \bibinfo {pages}
  {054009} (\bibinfo {year} {2018}{\natexlab{b}})}\BibitemShut {NoStop}%
\bibitem [{\citenamefont {Eckardt}(2017)}]{floquetgas}%
  \BibitemOpen
  \bibfield  {author} {\bibinfo {author} {\bibfnamefont {A.}~\bibnamefont
  {Eckardt}},\ }\bibfield  {title} {\emph {\bibinfo {title} {Colloquium: Atomic
  quantum gases in periodically driven optical lattices},\ }}\href {\doibase
  10.1103/RevModPhys.89.011004} {\bibfield  {journal} {\bibinfo  {journal}
  {Rev. Mod. Phys.}\ }\textbf {\bibinfo {volume} {89}},\ \bibinfo {pages}
  {011004} (\bibinfo {year} {2017})}\BibitemShut {NoStop}%
\bibitem [{\citenamefont {Weitenberg}\ and\ \citenamefont
  {Simonet}(2021)}]{floquetgas2}%
  \BibitemOpen
  \bibfield  {author} {\bibinfo {author} {\bibfnamefont {C.}~\bibnamefont
  {Weitenberg}}\ and\ \bibinfo {author} {\bibfnamefont {J.}~\bibnamefont
  {Simonet}},\ }\bibfield  {title} {\emph {\bibinfo {title} {Tailoring quantum
  gases by floquet engineering},\ }}\href {\doibase s41567-021-01316-x}
  {\bibfield  {journal} {\bibinfo  {journal} {Nat. Phys.}\ }\textbf {\bibinfo
  {volume} {17}},\ \bibinfo {pages} {1342} (\bibinfo {year}
  {2021})}\BibitemShut {NoStop}%
\bibitem [{\citenamefont {Garziano}\ \emph {et~al.}(2016)\citenamefont
  {Garziano}, \citenamefont {Macr\`{\i}}, \citenamefont {Stassi}, \citenamefont
  {Di~Stefano}, \citenamefont {Nori},\ and\ \citenamefont
  {Savasta}}]{secondorder}%
  \BibitemOpen
  \bibfield  {author} {\bibinfo {author} {\bibfnamefont {L.}~\bibnamefont
  {Garziano}}, \bibinfo {author} {\bibfnamefont {V.}~\bibnamefont
  {Macr\`{\i}}}, \bibinfo {author} {\bibfnamefont {R.}~\bibnamefont {Stassi}},
  \bibinfo {author} {\bibfnamefont {O.}~\bibnamefont {Di~Stefano}}, \bibinfo
  {author} {\bibfnamefont {F.}~\bibnamefont {Nori}}, \ and\ \bibinfo {author}
  {\bibfnamefont {S.}~\bibnamefont {Savasta}},\ }\bibfield  {title} {\emph
  {\bibinfo {title} {One photon can simultaneously excite two or more atoms},\
  }}\href {\doibase 10.1103/PhysRevLett.117.043601} {\bibfield  {journal}
  {\bibinfo  {journal} {Phys. Rev. Lett.}\ }\textbf {\bibinfo {volume} {117}},\
  \bibinfo {pages} {043601} (\bibinfo {year} {2016})}\BibitemShut {NoStop}%
\bibitem [{\citenamefont {Ma}\ and\ \citenamefont {Law}(2015)}]{ae}%
  \BibitemOpen
  \bibfield  {author} {\bibinfo {author} {\bibfnamefont {K.~K.~W.}\
  \bibnamefont {Ma}}\ and\ \bibinfo {author} {\bibfnamefont {C.~K.}\
  \bibnamefont {Law}},\ }\bibfield  {title} {\emph {\bibinfo {title}
  {Three-photon resonance and adiabatic passage in the large-detuning rabi
  model},\ }}\href {\doibase 10.1103/PhysRevA.92.023842} {\bibfield  {journal}
  {\bibinfo  {journal} {Phys. Rev. A}\ }\textbf {\bibinfo {volume} {92}},\
  \bibinfo {pages} {023842} (\bibinfo {year} {2015})}\BibitemShut {NoStop}%
\bibitem [{\citenamefont {Kaufman}\ \emph {et~al.}(2020)\citenamefont
  {Kaufman}, \citenamefont {Rozgonyi}, \citenamefont {Marquetand},\ and\
  \citenamefont {Weinacht}}]{ae1}%
  \BibitemOpen
  \bibfield  {author} {\bibinfo {author} {\bibfnamefont {B.}~\bibnamefont
  {Kaufman}}, \bibinfo {author} {\bibfnamefont {T.}~\bibnamefont {Rozgonyi}},
  \bibinfo {author} {\bibfnamefont {P.}~\bibnamefont {Marquetand}}, \ and\
  \bibinfo {author} {\bibfnamefont {T.}~\bibnamefont {Weinacht}},\ }\bibfield
  {title} {\emph {\bibinfo {title} {Adiabatic elimination in strong-field
  light-matter coupling},\ }}\href {\doibase 10.1103/PhysRevA.102.063117}
  {\bibfield  {journal} {\bibinfo  {journal} {Phys. Rev. A}\ }\textbf {\bibinfo
  {volume} {102}},\ \bibinfo {pages} {063117} (\bibinfo {year}
  {2020})}\BibitemShut {NoStop}%
\bibitem [{\citenamefont {Combescot}(2001)}]{fermigolden}%
  \BibitemOpen
  \bibfield  {author} {\bibinfo {author} {\bibfnamefont {M.}~\bibnamefont
  {Combescot}},\ }\bibfield  {title} {\emph {\bibinfo {title} {On the
  generalized golden rule for transition probabilities},\ }}\href {\doibase
  10.1088/0305-4470/34/31/304} {\bibfield  {journal} {\bibinfo  {journal} {J.
  Phys. A: Math. Gen.}\ }\textbf {\bibinfo {volume} {34}},\ \bibinfo {pages}
  {6087} (\bibinfo {year} {2001})}\BibitemShut {NoStop}%
\bibitem [{\citenamefont {Koch}\ \emph {et~al.}(2010)\citenamefont {Koch},
  \citenamefont {Houck}, \citenamefont {Hur},\ and\ \citenamefont
  {Girvin}}]{timereversal}%
  \BibitemOpen
  \bibfield  {author} {\bibinfo {author} {\bibfnamefont {J.}~\bibnamefont
  {Koch}}, \bibinfo {author} {\bibfnamefont {A.~A.}\ \bibnamefont {Houck}},
  \bibinfo {author} {\bibfnamefont {K.~L.}\ \bibnamefont {Hur}}, \ and\
  \bibinfo {author} {\bibfnamefont {S.~M.}\ \bibnamefont {Girvin}},\ }\bibfield
   {title} {\emph {\bibinfo {title} {Time-reversal-symmetry breaking in
  circuit-qed-based photon lattices},\ }}\href {\doibase
  10.1103/PhysRevA.82.043811} {\bibfield  {journal} {\bibinfo  {journal} {Phys.
  Rev. A}\ }\textbf {\bibinfo {volume} {82}},\ \bibinfo {pages} {043811}
  (\bibinfo {year} {2010})}\BibitemShut {NoStop}%
\bibitem [{\citenamefont {Xie}\ \emph {et~al.}(2020)\citenamefont {Xie},
  \citenamefont {Ma},\ and\ \citenamefont {Li}}]{Fock1}%
  \BibitemOpen
  \bibfield  {author} {\bibinfo {author} {\bibfnamefont {J.-k.}\ \bibnamefont
  {Xie}}, \bibinfo {author} {\bibfnamefont {S.-l.}\ \bibnamefont {Ma}}, \ and\
  \bibinfo {author} {\bibfnamefont {F.-l.}\ \bibnamefont {Li}},\ }\bibfield
  {title} {\emph {\bibinfo {title} {Quantum-interference-enhanced magnon
  blockade in an yttrium-iron-garnet sphere coupled to superconducting
  circuits},\ }}\href {\doibase 10.1103/PhysRevA.101.042331} {\bibfield
  {journal} {\bibinfo  {journal} {Phys. Rev. A}\ }\textbf {\bibinfo {volume}
  {101}},\ \bibinfo {pages} {042331} (\bibinfo {year} {2020})}\BibitemShut
  {NoStop}%
\bibitem [{\citenamefont {Liu}\ \emph {et~al.}(2019)\citenamefont {Liu},
  \citenamefont {Xiong},\ and\ \citenamefont {Wu}}]{Fock2}%
  \BibitemOpen
  \bibfield  {author} {\bibinfo {author} {\bibfnamefont {Z.-X.}\ \bibnamefont
  {Liu}}, \bibinfo {author} {\bibfnamefont {H.}~\bibnamefont {Xiong}}, \ and\
  \bibinfo {author} {\bibfnamefont {Y.}~\bibnamefont {Wu}},\ }\bibfield
  {title} {\emph {\bibinfo {title} {Magnon blockade in a hybrid
  ferromagnet-superconductor quantum system},\ }}\href {\doibase
  10.1103/PhysRevB.100.134421} {\bibfield  {journal} {\bibinfo  {journal}
  {Phys. Rev. B}\ }\textbf {\bibinfo {volume} {100}},\ \bibinfo {pages}
  {134421} (\bibinfo {year} {2019})}\BibitemShut {NoStop}%
\bibitem [{\citenamefont {Kounalakis}\ \emph {et~al.}(2022)\citenamefont
  {Kounalakis}, \citenamefont {Bauer},\ and\ \citenamefont
  {Blanter}}]{catstate}%
  \BibitemOpen
  \bibfield  {author} {\bibinfo {author} {\bibfnamefont {M.}~\bibnamefont
  {Kounalakis}}, \bibinfo {author} {\bibfnamefont {G.~E.~W.}\ \bibnamefont
  {Bauer}}, \ and\ \bibinfo {author} {\bibfnamefont {Y.~M.}\ \bibnamefont
  {Blanter}},\ }\bibfield  {title} {\emph {\bibinfo {title} {Analog quantum
  control of magnonic cat states on a chip by a superconducting qubit},\
  }}\href {\doibase 10.1103/PhysRevLett.129.037205} {\bibfield  {journal}
  {\bibinfo  {journal} {Phys. Rev. Lett.}\ }\textbf {\bibinfo {volume} {129}},\
  \bibinfo {pages} {037205} (\bibinfo {year} {2022})}\BibitemShut {NoStop}%
\bibitem [{\citenamefont {Wang}\ \emph
  {et~al.}(2016{\natexlab{b}})\citenamefont {Wang}, \citenamefont {Zhang},
  \citenamefont {Zhang}, \citenamefont {Luo}, \citenamefont {Xiong},
  \citenamefont {Wang}, \citenamefont {Li}, \citenamefont {Hu},\ and\
  \citenamefont {You}}]{kerrmagnon2}%
  \BibitemOpen
  \bibfield  {author} {\bibinfo {author} {\bibfnamefont {Y.-P.}\ \bibnamefont
  {Wang}}, \bibinfo {author} {\bibfnamefont {G.-Q.}\ \bibnamefont {Zhang}},
  \bibinfo {author} {\bibfnamefont {D.}~\bibnamefont {Zhang}}, \bibinfo
  {author} {\bibfnamefont {X.-Q.}\ \bibnamefont {Luo}}, \bibinfo {author}
  {\bibfnamefont {W.}~\bibnamefont {Xiong}}, \bibinfo {author} {\bibfnamefont
  {S.-P.}\ \bibnamefont {Wang}}, \bibinfo {author} {\bibfnamefont {T.-F.}\
  \bibnamefont {Li}}, \bibinfo {author} {\bibfnamefont {C.-M.}\ \bibnamefont
  {Hu}}, \ and\ \bibinfo {author} {\bibfnamefont {J.~Q.}\ \bibnamefont {You}},\
  }\bibfield  {title} {\emph {\bibinfo {title} {Magnon kerr effect in a
  strongly coupled cavity-magnon system},\ }}\href {\doibase
  10.1103/PhysRevB.94.224410} {\bibfield  {journal} {\bibinfo  {journal} {Phys.
  Rev. B}\ }\textbf {\bibinfo {volume} {94}},\ \bibinfo {pages} {224410}
  (\bibinfo {year} {2016}{\natexlab{b}})}\BibitemShut {NoStop}%
\bibitem [{\citenamefont {Qi}\ and\ \citenamefont
  {Jing}(2022{\natexlab{b}})}]{population}%
  \BibitemOpen
  \bibfield  {author} {\bibinfo {author} {\bibfnamefont {S.-f.}\ \bibnamefont
  {Qi}}\ and\ \bibinfo {author} {\bibfnamefont {J.}~\bibnamefont {Jing}},\
  }\bibfield  {title} {\emph {\bibinfo {title} {Accelerated adiabatic passage
  in cavity magnomechanics},\ }}\href {\doibase 10.1103/PhysRevA.105.053710}
  {\bibfield  {journal} {\bibinfo  {journal} {Phys. Rev. A}\ }\textbf {\bibinfo
  {volume} {105}},\ \bibinfo {pages} {053710} (\bibinfo {year}
  {2022}{\natexlab{b}})}\BibitemShut {NoStop}%
\bibitem [{\citenamefont {Wootters}(1998)}]{concurrence}%
  \BibitemOpen
  \bibfield  {author} {\bibinfo {author} {\bibfnamefont {W.~K.}\ \bibnamefont
  {Wootters}},\ }\bibfield  {title} {\emph {\bibinfo {title} {Entanglement of
  formation of an arbitrary state of two qubits},\ }}\href {\doibase
  10.1103/PhysRevLett.80.2245} {\bibfield  {journal} {\bibinfo  {journal}
  {Phys. Rev. Lett.}\ }\textbf {\bibinfo {volume} {80}},\ \bibinfo {pages}
  {2245} (\bibinfo {year} {1998})}\BibitemShut {NoStop}%
\bibitem [{\citenamefont {Breuer}\ and\ \citenamefont
  {Petruccione}(2022)}]{opentheory}%
  \BibitemOpen
  \bibfield  {author} {\bibinfo {author} {\bibfnamefont {H.-P.}\ \bibnamefont
  {Breuer}}\ and\ \bibinfo {author} {\bibfnamefont {F.}~\bibnamefont
  {Petruccione}},\ }\bibfield  {title} {\emph {\bibinfo {title} {The theory of
  open quantum systems},\ }}\href@noop {} {\bibfield  {journal} {\bibinfo
  {journal} {Oxford University Press, Oxford}\ } (\bibinfo {year}
  {2022})}\BibitemShut {NoStop}%
\bibitem [{\citenamefont {Viola}\ and\ \citenamefont {Lloyd}(1998)}]{suppre}%
  \BibitemOpen
  \bibfield  {author} {\bibinfo {author} {\bibfnamefont {L.}~\bibnamefont
  {Viola}}\ and\ \bibinfo {author} {\bibfnamefont {S.}~\bibnamefont {Lloyd}},\
  }\bibfield  {title} {\emph {\bibinfo {title} {Dynamical suppression of
  decoherence in two-state quantum systems},\ }}\href {\doibase
  10.1103/PhysRevA.58.2733} {\bibfield  {journal} {\bibinfo  {journal} {Phys.
  Rev. A}\ }\textbf {\bibinfo {volume} {58}},\ \bibinfo {pages} {2733}
  (\bibinfo {year} {1998})}\BibitemShut {NoStop}%
\bibitem [{\citenamefont {Niemczyk}\ \emph {et~al.}(2010)\citenamefont
  {Niemczyk}, \citenamefont {Deppe}, \citenamefont {Huebl}, \citenamefont
  {Menzel}, \citenamefont {Hocke}, \citenamefont {Schwarz}, \citenamefont
  {Garciaripoll}, \citenamefont {Zueco}, \citenamefont {H\"ommer},\ and\
  \citenamefont {Solano}}]{cq}%
  \BibitemOpen
  \bibfield  {author} {\bibinfo {author} {\bibfnamefont {T.}~\bibnamefont
  {Niemczyk}}, \bibinfo {author} {\bibfnamefont {F.}~\bibnamefont {Deppe}},
  \bibinfo {author} {\bibfnamefont {H.}~\bibnamefont {Huebl}}, \bibinfo
  {author} {\bibfnamefont {E.~P.}\ \bibnamefont {Menzel}}, \bibinfo {author}
  {\bibfnamefont {F.}~\bibnamefont {Hocke}}, \bibinfo {author} {\bibfnamefont
  {M.~J.}\ \bibnamefont {Schwarz}}, \bibinfo {author} {\bibfnamefont {J.~J.}\
  \bibnamefont {Garciaripoll}}, \bibinfo {author} {\bibfnamefont
  {D.}~\bibnamefont {Zueco}}, \bibinfo {author} {\bibfnamefont
  {T.}~\bibnamefont {H\"ommer}}, \ and\ \bibinfo {author} {\bibfnamefont
  {E.}~\bibnamefont {Solano}},\ }\bibfield  {title} {\emph {\bibinfo {title}
  {Circuit quantum electrodynamics in the ultrastrong-coupling regime},\
  }}\href {https://www.nature.com/articles/nphys1730} {\bibfield  {journal}
  {\bibinfo  {journal} {Nat. Phys.}\ }\textbf {\bibinfo {volume} {6}},\
  \bibinfo {pages} {772} (\bibinfo {year} {2010})}\BibitemShut {NoStop}%
\bibitem [{\citenamefont {Rudner}\ \emph {et~al.}(2013)\citenamefont {Rudner},
  \citenamefont {Lindner}, \citenamefont {Berg},\ and\ \citenamefont
  {Levin}}]{edgestate}%
  \BibitemOpen
  \bibfield  {author} {\bibinfo {author} {\bibfnamefont {M.~S.}\ \bibnamefont
  {Rudner}}, \bibinfo {author} {\bibfnamefont {N.~H.}\ \bibnamefont {Lindner}},
  \bibinfo {author} {\bibfnamefont {E.}~\bibnamefont {Berg}}, \ and\ \bibinfo
  {author} {\bibfnamefont {M.}~\bibnamefont {Levin}},\ }\bibfield  {title}
  {\emph {\bibinfo {title} {Anomalous edge states and the bulk-edge
  correspondence for periodically driven two-dimensional systems},\ }}\href
  {\doibase 10.1103/PhysRevX.3.031005} {\bibfield  {journal} {\bibinfo
  {journal} {Phys. Rev. X}\ }\textbf {\bibinfo {volume} {3}},\ \bibinfo {pages}
  {031005} (\bibinfo {year} {2013})}\BibitemShut {NoStop}%
\end{thebibliography}%

\end{document}